\documentclass[%
reprint,
 amsmath,amssymb,
 aps,
]{revtex4-1}

\usepackage{graphicx}
\usepackage{dcolumn}
\usepackage{bm}
\usepackage{hyperref}

\usepackage{afterpage}

\begin{document}

\title{Electron Accelerator for Radiation Therapy with Beam Energy 6 -- 20~MeV}

\author{L.Yu.~Ovchinnikova}
\email{lub.ovch@yandex.ru}
\author{V.I.~Shvedunov}
\affiliation{%
 Skobeltsyn Institute of Nuclear Physics, Lomonosov Moscow State University (SINP MSU), Moscow 119991, Russia \\
 Laboratory of Electron Accelerators MSU Ltd (LEA MSU), Moscow 119234, Russia
}%

\date{\today}

\begin{abstract}
\begin{description}
	\item[Purpose]
	To describe a concept of a compact electron accelerator for external radiation therapy with variable energy in the range of 6 -- 20~MeV, based on linotron principle.
	\item[Methods]
	Beam dynamics simulation using the CST and MAD-X code.
	Various optimization methods of multi-parameter problem.
	\item[Results]
	Our accelerator differs from the Reflexotron in a number of essential details:
	a much more compact and more efficient C-band accelerating structure, optimized for the high capture efficiency, narrow energy and phase spectra, and low transverse emittance;
	magnetic mirror with fixed field based on rare-earth permanent magnets;
	three-electrode electron gun with off-axis placement of the cathode with a current regulated in the range of two orders of magnitude.
	These improvements allow the possibility to:
	adjust the accelerated beam current in a wide range in accordance with the required energy;
	reduce parasitic losses of the beam current and the associated parasitic radiation;
	eliminate the risk of setting an erroneous energy value;
	significantly reduce the dimensions of the accelerator and simplify its operation.
	\item[Conclusions]
	We presented the results of calculation of the electron accelerator for external radiation therapy in the energy range of 6 -- 20~MeV. The accelerator is based on the principle of double beam acceleration in the same accelerating structure, which allows to control the beam energy in a wide range, reduce RF power consumption and the dimensions of the accelerator, and, therefore, reduce its cost. The results can be used to develop the design of the accelerator on the platform of the KLT-6 complex created by ROSATOM.
\end{description}
Keywords: electron accelerator, magnetic mirror, radiation therapy
\end{abstract}

\maketitle

\tableofcontents

\section{Introduction}

The principles of construction of electron accelerators for radiation therapy are described in detail in the literature \cite{Karzmark__1993__Medical_Electron_Accelerators}.
Modern linear accelerators for external radiation therapy with low-energy bremsstrahlung (6~MeV) have a relatively simple and compact sealed-off accelerating system, including an accelerating structure, an electron gun, a bremsstrahlung target, an RF vacuum window and an ion pump.
A compact source of RF energy - magnetron - is located near the accelerating system.
A magnetic system is not required to form an irradiation field.
This allows a fairly wide range of manufacturers to develop their own complexes of radiation therapy, not only with the classic isocentric gantry, but also with other equipment placement schemes and the organization of the irradiation process, such as Cyberknife \cite{Adler__1997__The_Cyberknife_A_Frameless_Robotic_System_for_Radiosurgery}, tomotherapy \cite{Mackie__2006__History_of_tomotherapy}, combining the accelerator with MRI \cite{Liney__2018__MRI_Linear_Accelerator_Radiotherapy_Systems}.

The authors of this work took part in the development of a 6~MeV linear accelerator for the KLT-6 radiation therapy complex designed by ROSATOM \cite{Rodko__2019__Development_of_a_Radiotherapy_System_Based_on_6_MeV_Linac_and_Cone-Beam_Computer_Tomograph}.
The linear accelerator is based on the C-band accelerating structure, 225 mm long, powered by a compact multi-beam klystron with a focusing system based on rare-earth permanent magnets with a maximum RF power of 3.6~MW \cite{Ovchinnikova__2018__Calculation_of_Electron-Beam_Dynamics_in_a_C_Band_Accelerator_for_a_Radiotherapy_Complex, Yurov__2019__Beam_Parameters_Measurement_of_C_band_6_MeV_Linear_Electron_Accelerator}.
In the therapeutic dose providing mode, the accelerator produces a maximum dose rate of 10~Gy/min without a flattering filter at a distance of 1~m.
In the portal image acquisition mode, in order to increase contrast and reduce the patient dose, the accelerated beam energy is reduced to 2.5~MeV.

Different situation takes place with medium and high energy linear electron accelerators for external radiation therapy (18 -- 25~MeV).
Currently, the vast majority of such accelerators are manufactured by Varian Medical Systems \cite{Varian} and Elekta \cite{Elekta}. These accelerators are being improved to expand their functionality, increase accuracy, stability, reliability, safety, and develop new treatment methods \cite{Wilson__2015__Industrial_Design}.
The experience gained by these companies over decades makes it difficult to create competitive analogues based on the same principles by other manufacturers.

At the same time, using other principles, it is possible to design cheaper and more compact electron accelerators for radiation therapy at medium and high energies. In the late 70's. of the last century, the Atomic Energy of Canada Limited developed original electron accelerator for radiation therapy with an energy of 5 -- 25~MeV - a reflexotron \cite{Schriber__1975__Double_Pass_Linear_Accelerator_Reflexotron, Schriber__1977__Experimental_Measurements_on_a_25_Mev_Reflexotron, Taylor__1983__Therac_25_A_New_Medical_Accelerator_Concept}.
It was based on the principle of the linotron \cite{Kolomensky__1967__Linotron_A_Proposed_System_Off_Particle_Acceleration} - acceleration of electrons in the forward and reverse directions in a standing wave accelerating structure using a magnetic mirror to return the beam to the accelerating structure.
This principle makes it possible to realize a compact accelerator with low RF power consumption, with the possibility to vary beam energy in a wide range.
However, for a number of reasons, including the low dose rate at low energy, complexity and high cost of production \cite{Karzmark__1993__Medical_Electron_Accelerators}, problems with the control system, the reflexotron was not widely used and its production was discontinued.

Development of 3D methods of computer simulation of accelerating structures, electrostatic and magnetostatic systems and beam dynamics, which allow with high accuracy to calculate and to design accelerators components; the use a new RF band in which compact and powerful klystrons appeared; the development of accelerator control systems, make it possible to reconsider possibility to build medical electron accelerator based on linotron principle.

This paper presents the results of a full 3D computer simulation of electron accelerator for external radiation therapy for energy, regulated in the range of 6 -- 20~MeV, based on linotron principle.
Due to high efficiency of RF energy utilization, the accelerator can be built on the 6~MeV accelerator platform of the KLT-6 complex \cite{Rodko__2019__Development_of_a_Radiotherapy_System_Based_on_6_MeV_Linac_and_Cone-Beam_Computer_Tomograph}, i.e. using all its systems, with the exception of the accelerating structure and the electron gun, and with the addition of the necessary magnetic systems.

Our approach to design of accelerator with a magnetic mirror differs from \cite{Schriber__1975__Double_Pass_Linear_Accelerator_Reflexotron, Schriber__1977__Experimental_Measurements_on_a_25_Mev_Reflexotron, Taylor__1983__Therac_25_A_New_Medical_Accelerator_Concept} in a number of essential details: a much more compact and more efficient C-band accelerating structure, optimized for the high capture efficiency, the narrow energy and phase spectra, and low transverse emittance; magnetic mirror with fixed field based on rare-earth permanent magnets; three-electrode electron gun with off-axis placement of the cathode with a current regulated in the range of two orders of magnitude.
These changes make it possible to adjust the accelerated beam current in a wide range in accordance with the set energy; to reduce parasitic losses of the beam current and the associated parasitic radiation; eliminate the possibility of setting an erroneous energy value; significantly reduce the size of the accelerator and simplify its operation.

In carrying out this work, in addition to the experience that we obtained in developing a linear accelerator for the KLT-6 complex, we also used our experience in building electron accelerator with magnetic mirror \cite{Alimov__1998__Magnetic_Mirror_for_a_Continuous_Wave_Electron_Accelerator}; race-track microtrons, including with magnetic systems based on rare-earth permanent magnets \cite{Shvedunov__2004__A_racetrack_microtron_with_high_brightness_beams, Shvedunov__2005__A_70_Mev_racetrack_microtron, Ermakov__2018__A_Multipurpose_Pulse_Race_Track_Microtron_with_an_Energy_of_55_MeV, Kubyshin__2010__Current_Status_of_the_12_MeV_UPC_Race_Track_Microtron}, in which, in some cases, the principle of reflection and re-acceleration of the beam is also used; the experience of developing electron guns with off-axis placement of the cathode and with the electrostatic method of guiding the beam to the axis of accelerating structure \cite{Aloev__2010__Electron_gun_with_off_axis_beam_injection_for_a_race_track_microtron, Ovchinnikova__2015__Optimization_of_an_electron_gun_with_an_off_axis_cathode_position}; as well as the experience of building applied linear electron accelerators for various purposes \cite{Shvedunov__2019__Electron_accelerators_design_and_construction_at_Lomonosov_Moscow_State_University}.

\section{%
	\label{sec:Accelerator_scheme}%
	Accelerator scheme}

\begin{figure}[thbp]
	\begin{minipage}[t]{1\linewidth}
		\begin{center}
			\includegraphics[width=1\linewidth]{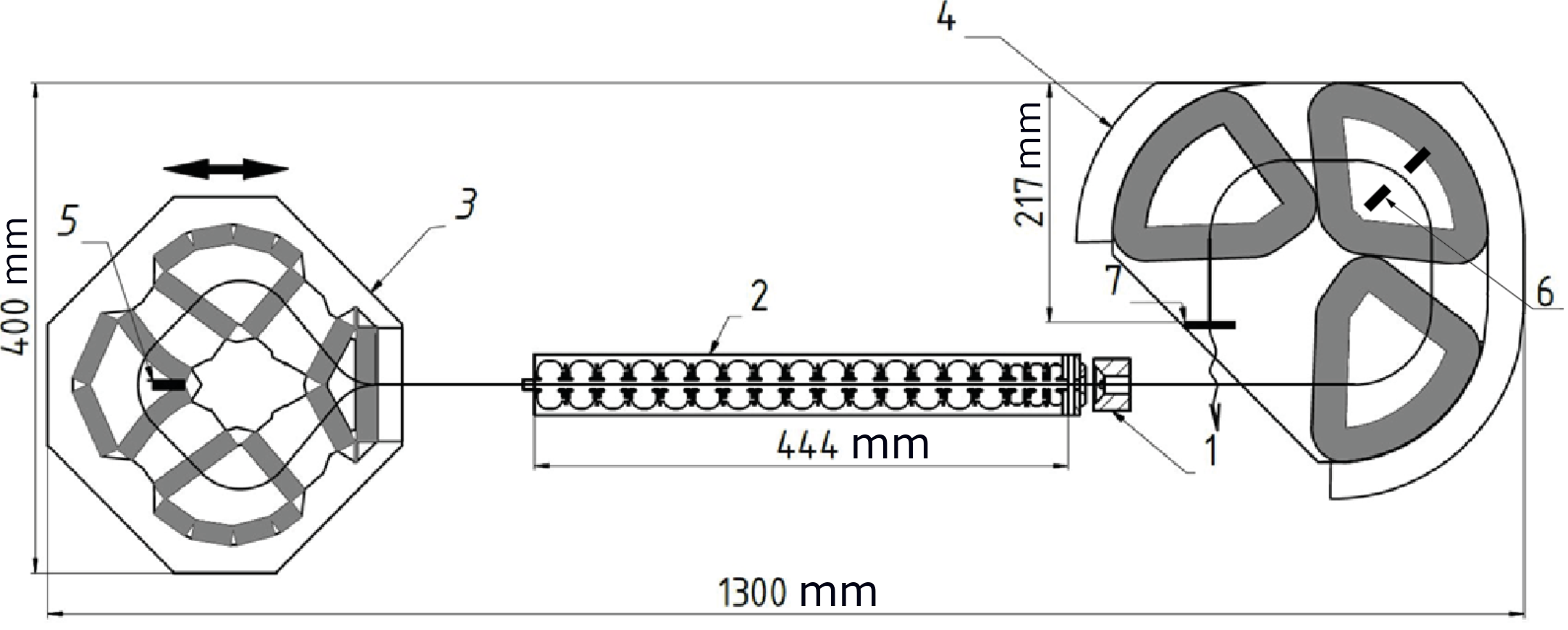}\\
			(a)
		\end{center}
	\end{minipage}
	\begin{minipage}[t]{0.49\linewidth}
		\begin{center}
			\includegraphics[width=1\linewidth]{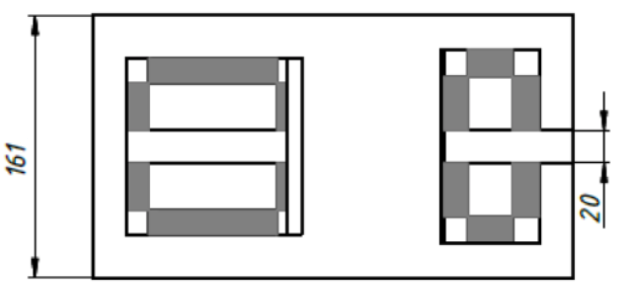}\\
			(b)
		\end{center}
	\end{minipage}
	\hfill
	\begin{minipage}[t]{0.49\linewidth}
		\begin{center}
			\includegraphics[width=1\linewidth]{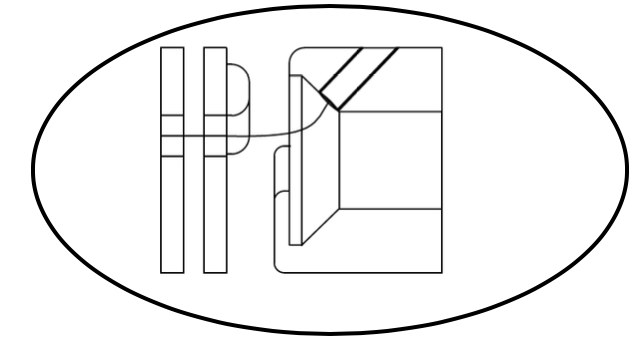}\\
			(c)
		\end{center}
	\end{minipage}
	\caption{%
		\label{fig:accelerator_scheme}%
		(a) Accelerator scheme.
		\textit{1} --- electron gun,
		\textit{2} --- accelerating structure,
		\textit{3} --- magnetic mirror,
		\textit{4} --- $270^{\circ}$  bending magnet,
		\textit{5}, \textit{6} --- collimators,
		\textit{7} --- bremsstrahlung target.
		(b, c) --- sections in the perpendicular plane, respectively, of the magnetic mirror and electron gun.
	}
\end{figure}

The accelerator scheme is shown in Fig.~\ref{fig:accelerator_scheme}. Its main elements, the optimization of which was carried out in the framework of this work, are the electron gun, accelerating structure, magnetic mirror, and $270^\circ$ bending magnet.

\begin{figure}[thbp]
	\begin{minipage}[t]{0.49\linewidth}
		\begin{center}
			\includegraphics[width=1\linewidth]{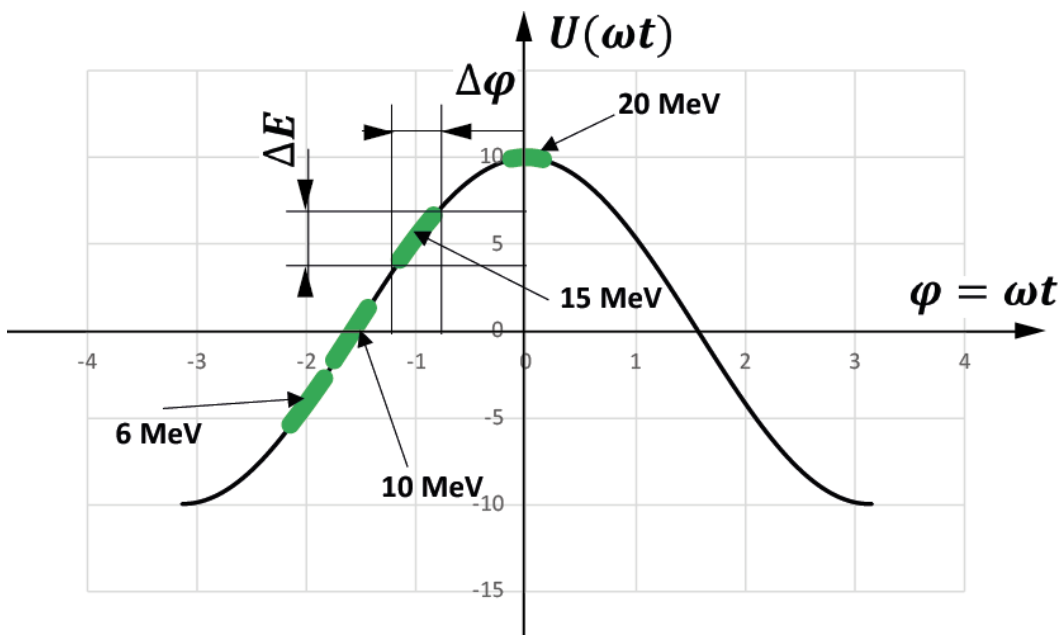}\\
			(a)
		\end{center}
	\end{minipage}
	\hfill
	\begin{minipage}[t]{0.49\linewidth}
		\begin{center}
			\includegraphics[width=1\linewidth]{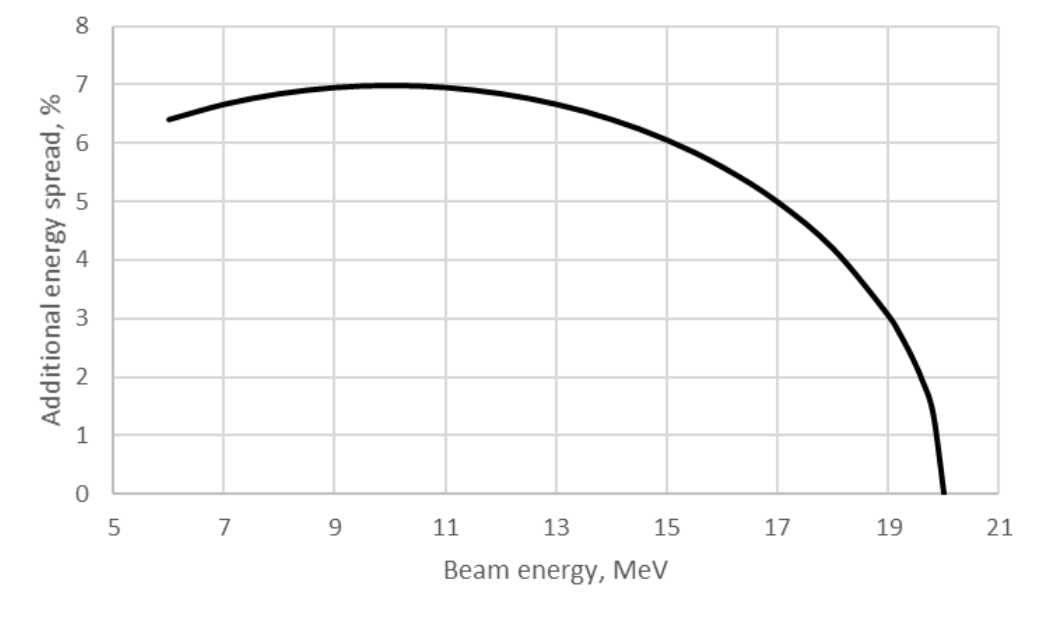}\\
			(b)
		\end{center}
	\end{minipage}
	\caption{%
		\label{fig:phase}%
		(a) The position of the bunch relative to the phase of the accelerating field for various final energies.
		(b) Additional relative energy spread for the phase length of the bunch $4^{\circ}$.
	}
\end{figure}

The accelerator operates as follows.
The electron beam from the gun \textit{1} with off-axis placement of the cathode is injected into the accelerating structure \textit{2} and accelerated to the energy of 10~MeV.
After passing through the magnetic mirror \textit{3}, the beam returns to the accelerating structure and is accelerated or decelerated in it to a set energy, depending on the input phase determined by the distance between the structure and the mirror.
After passing through the holes in the gun’s electrodes, the beam enters the magnet \textit{4}, which bends it for $270^{\circ}$, directing it either to the bremsstrahlung target \textit{7} or to the exit window, depending on the selected operating mode for the radiation used.

The operation of a linear accelerator in the linotron mode with regulation of the final energy due to a change in the phase of the bunch relative to the phase of the accelerating field has a feature illustrated in Fig.~\ref{fig:phase}~(a).
In the reverse motion in some phase $\phi$, a bunch of length $\Delta $$\phi$, which initially had zero energy spread, will have a relative energy spread at the exit:%
\begin{equation}
\frac{\Delta E}{E_b} = 2 \sin(\phi) \sin(\frac{\Delta \phi}{2})\approx  sin(\phi) \Delta \phi .
\label{eq:energy_spread}
\end{equation}

The maximum relative energy spread is achieved in the phase
$\phi = \pi / 2 : \Delta E / E_b = \Delta \phi$.
The dependence of the relative energy spread on energy for a bunch with a phase length of $4^{\circ}$, which initially had zero energy spread, is shown in Fig.~\ref{fig:phase}~(b).
The maximum relative spread of 7\% is achieved at an energy of 10~MeV.
In the phase of maximum acceleration, the beam does not acquire additional energy spread.
Thus, with a fixed increase in energy in the accelerating structure during the first passage, to reduce the energy spread of the beam, the phase length of the bunch during second passage should be minimal.

To reduce the energy spread, the collimator \textit{5}, installed in magnetic mirror in the region with maximum dispersion, cuts off the low-energy electrons, allowing the formation of bunches of small phase length by means of mirror longitudinal dispersion.
In addition, a second collimator \textit{6} is installed in $270^{\circ}$  magnet, which, if necessary, can be used to further limit the width of the energy spectrum.

Another possibility of reducing the energy spread is the formation of bunches with such a correlation of the energy and phase in the longitudinal phase space, for which under second acceleration there is a partial compensation of the energy spread during acceleration in a nonzero phase.
In this case, the introduced energy spread will be preserved upon acceleration in the zero phase.
If we introduce the correlated relative energy spread
$(\Delta E/E_b)_i=\Delta \phi(|\sin(\phi)|-\alpha)$, where $\alpha<1$, then the final spread will be
$(\Delta E/E_b)_f=\Delta \phi \left| |\sin(\phi)|-\alpha \right|  /(|\cos(\phi)|+1)$.
For $\alpha \approx 0.67:(\Delta E/E_b)_f \approx 0.33\Delta \phi$, both for $\phi=\pi / 2$ and $\phi=0$.
Thus, for the phase length of the bunch $\pm2^{\circ}$, the relative spread will be $\pm2.3\%$.
The necessary correlation of the energy and phase of the particles can be achieved by choosing the parameters of a linear accelerator and a magnetic mirror.
However, the realization of this possibility is limited by nonlinear effects during the first acceleration of a bunch in a phase close to the phase of maximum energy gain.

\section{%
	\label{sec:Accelerator_System_Requirements}%
	Accelerator system requirements}

Assuming that the maximum average dose rate of bremsstrahlung at a distance of 1~m from the target in the whole energy range $E_b = 6 \dots 20 \text{~MeV}$, with the flattering filter for a field of $40\times40 \text{~cm}^2$, should be $D=5 \text{~Gy/min}$, we estimate the range of variation of the accelerated beam current.
Based on the data given in \cite{Karzmark__1993__Medical_Electron_Accelerators}~(p.~18), we obtain an approximate relationship between the pulsed beam current at the bremsstrahlung target and its energy:%
\begin{equation}
I_b\approx \frac{Dq}{1.6E_b^{1.8}} \text{~mA}
\label{eq:current_energy_relationship}
\end{equation}
where $q = T / \tau$ is off-duty factor of the accelerated beam current, $T$ --- repetition period, $\tau$ --- pulse duration.

\begin{figure}[thbp]
	\begin{center}
		\includegraphics[width=0.5\linewidth]{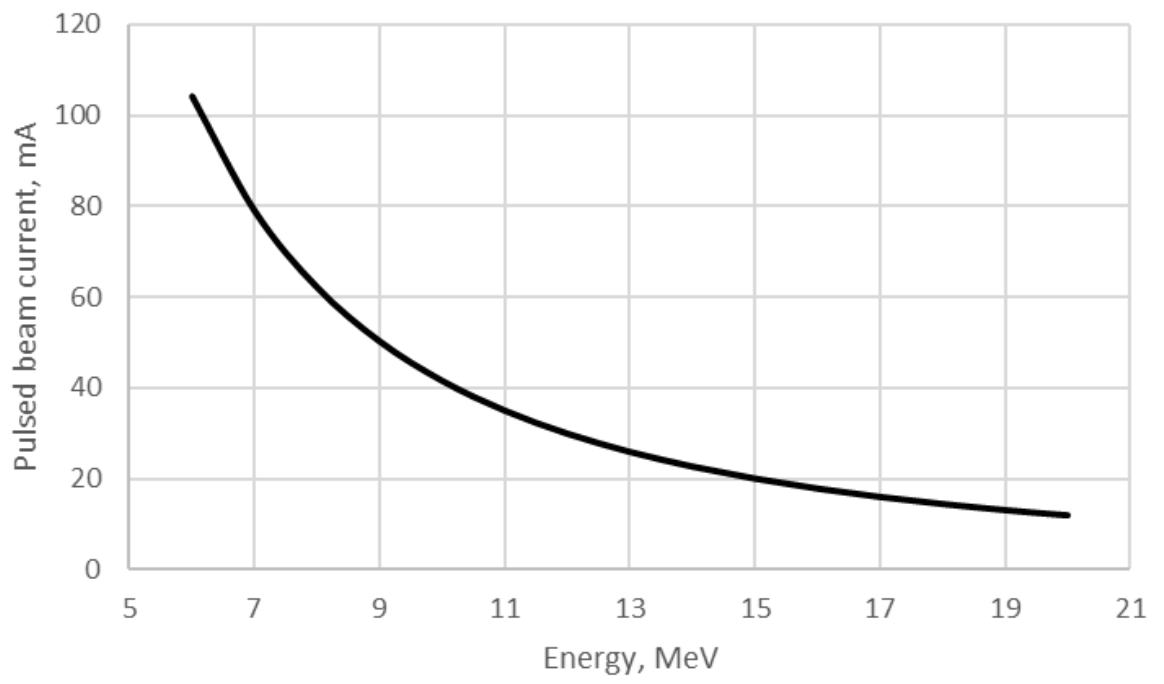}
		\caption{%
			\label{fig:current_energy_relationship}%
			The relationship of the pulse current and the energy of the accelerated beam.
		}
	\end{center}
\end{figure}

The relationship between the pulsed current and the accelerated beam energy, estimated by formula (\ref{eq:current_energy_relationship}) for the minimum off-duty factor $q = 840$ of the KLT-6 complex RF system, which can be used in this accelerator, is shown in Fig.~\ref{fig:current_energy_relationship}.
Preliminary calculations demonstrated that the total capture efficiency, taking into account the loss of particles at the collimators, can be about 30\%.
Thus, the pulse current of the gun for the bremsstrahlung generation mode should be regulated in the range from
$I_{gun}\approx 330 \dots 350 \text{~mA}$ at an energy of 6~MeV to
$I_{gun}\approx 40 \dots 50 \text{~mA}$ at an energy of 20~MeV.
Given estimation of the gun current are approximate; the actual choice of the injection current will be made during the accelerator calibration.

When operating in the electron beam mode, the average current should be approximately three orders of magnitude lower \cite{Karzmark__1993__Medical_Electron_Accelerators}~(p.~163).
Since obtaining acceptable characteristics of the gun beam at a maximum current of hundreds of milliamps for a current noticeably less than one milliampere is problematic, a decrease in the average current can be achieved by an additional increase in off-duty factor both by decrease of the repetition rate and the duration of the current pulse.

The parameters of the accelerating structure must satisfy the following requirements: high effective shunt impedance; the diameter of the beam channel, sufficient for the passage of the beam with low current loss; short length; the total RF power loss in the walls that provides beam energy after the first acceleration of 10~MeV, at a level of 2.5 -- 2.7~MW, so that the power of a 3.6~MW klystron is sufficient to accelerate a beam with a current of more than 100~mA at an energy of 6~MeV.
In addition, during the first acceleration of the beam to an energy of 10~MeV, the accelerating structure should provide a high capture efficiency of the electron gun current, a narrow energy spectrum, correlation of the energies and phases in the longitudinal phase space, should allow to compress the bunch with a magnetic mirror.

The basic requirements for a magnetic mirror are as follows: zero linear and angular dispersion; focusing properties that permit to accelerate or slow down the beam when moving through the accelerating structure in the opposite direction without the use of additional focusing lenses; providing compression of the bunch in the longitudinal phase space; small dimensions and weight.

As for the magnetic system that directs the accelerated beam to the bremsstrahlung target or to the exit window, today there are many options for constructing such a system \cite{Karzmark__1993__Medical_Electron_Accelerators}.
A magnetic system should have zero linear and angular dispersion and have focusing properties that provide optimal dimensions, divergence, and beam symmetry at the bremsstrahlung target.

\section{Electron gun}

Two versions of a two-electrode electron gun with off-axis cathode placement and electrostatic displacement of the beam to the axis of the accelerating structure described in \cite{Aloev__2010__Electron_gun_with_off_axis_beam_injection_for_a_race_track_microtron, Ovchinnikova__2015__Optimization_of_an_electron_gun_with_an_off_axis_cathode_position} differ in the maximum beam current.
Current control in these options is possible only by changing the voltage at the cathode. Such a control method leads to a change in the energy of the injected electrons and, as a consequence, to non-optimal conditions for the bunching and focusing of particles in the initial part of the structure.

\begin{figure}[thbp]
	\begin{minipage}[t]{0.49\linewidth}
		\begin{center}
			\includegraphics[width=1\linewidth]{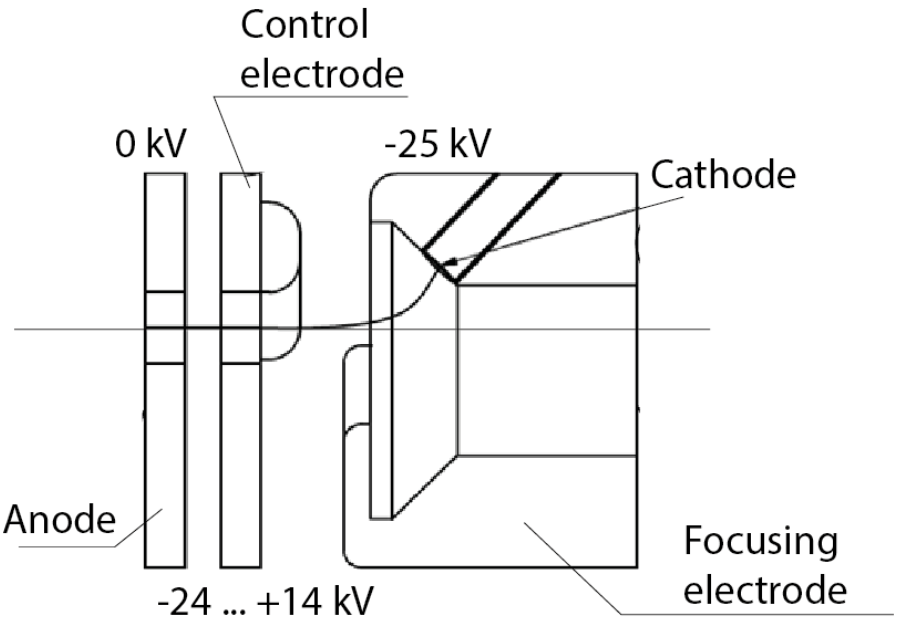}\\
			(a)
		\end{center}
	\end{minipage}
	\hfill
	\begin{minipage}[t]{0.49\linewidth}
		\begin{center}
			\includegraphics[width=1\linewidth]{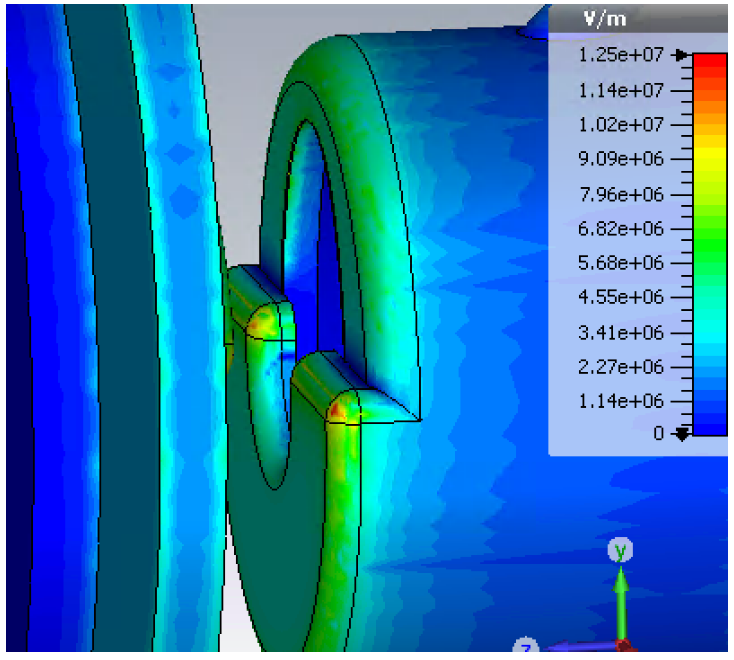}\\
			(b)
		\end{center}
	\end{minipage}
	\caption{%
		\label{fig:gun}%
		(a) Geometry of a three-electrode electron gun with off-axis cathode placement.
		(b) Distribution of electric field strength at the surface of electrodes.
	}
\end{figure}

In the present work, using the CST code \cite{CST}, a three-electrode gun was optimized with a voltage of -25~kV at the cathode and a current regulated in the range from 330~mA to about 1~mA by the change of the control electrode voltage relative to the anode in the range from -24~kV to +14~kV.
We used in calculations a spherical cathode with a diameter of 5~mm and a temperature of 1300~K.
The geometry of the optimized gun is shown in Fig.~\ref{fig:gun}.

\begin{figure}[thbp]
	\begin{minipage}[t]{0.49\linewidth}
		\begin{center}
			\includegraphics[width=1\linewidth]{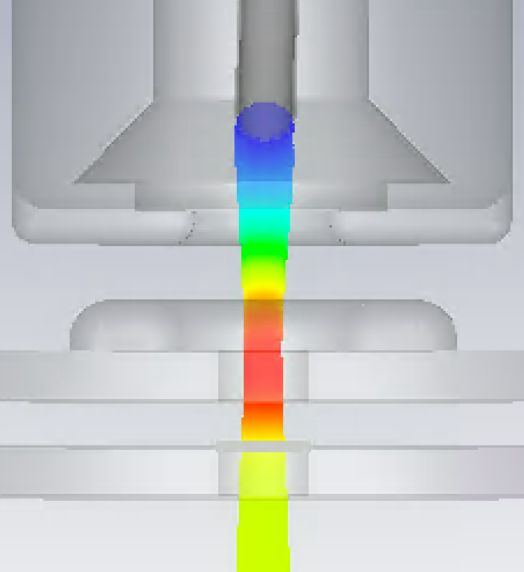}\\
			(a)
		\end{center}
	\end{minipage}
	\hfill
	\begin{minipage}[t]{0.49\linewidth}
		\begin{center}
			\includegraphics[width=1\linewidth]{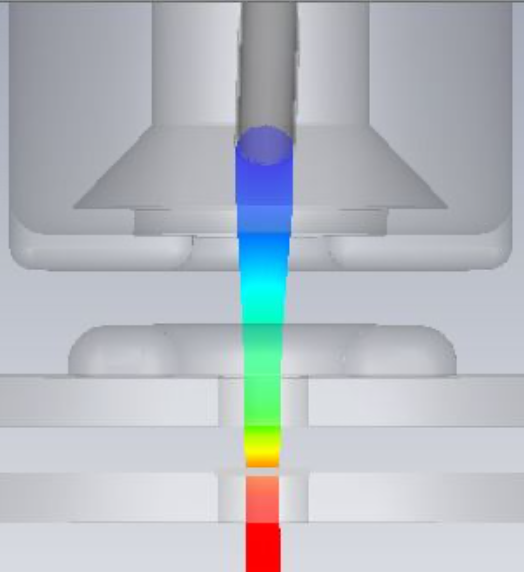}\\
			(b)
		\end{center}
	\end{minipage}
	\vspace{1pt}
	\begin{minipage}[t]{0.49\linewidth}
		\begin{center}
			\includegraphics[width=1\linewidth]{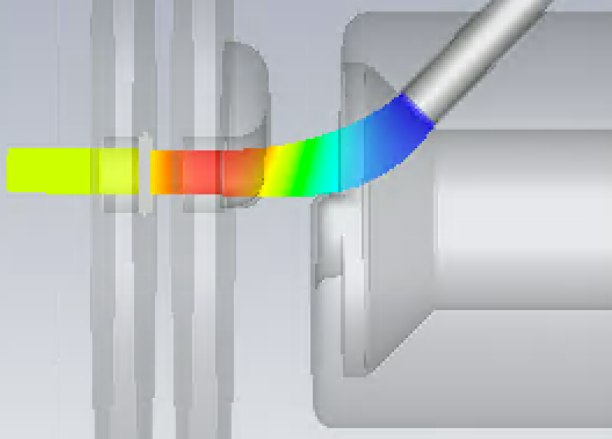}\\
			(c)
		\end{center}
	\end{minipage}
	\hfill
	\begin{minipage}[t]{0.49\linewidth}
		\begin{center}
			\includegraphics[width=1\linewidth]{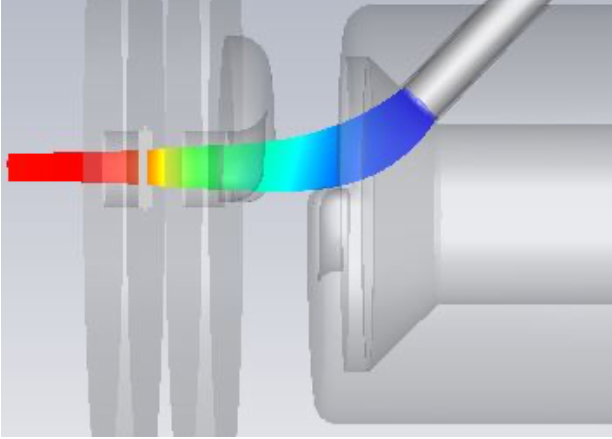}\\
			(d)
		\end{center}
	\end{minipage}
	\caption{%
		\label{fig:gun_trajectories}%
		Particle trajectories in two planes at voltages at the control electrode and beam current:
		(a, c): +13~kV (330~mA),
		(b, d): -14~kV (50~mA).
	}
\end{figure}

\begin{figure}[thbp]
	\begin{minipage}[t]{0.49\linewidth}
		\begin{center}
			\includegraphics[width=1\linewidth]{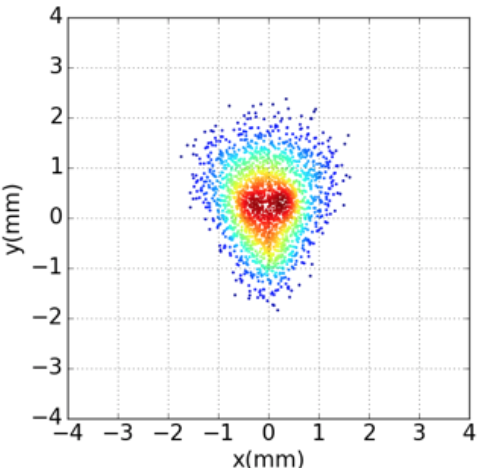}\\
			(a)
		\end{center}
	\end{minipage}
	\hfill
	\begin{minipage}[t]{0.49\linewidth}
		\begin{center}
			\includegraphics[width=1\linewidth]{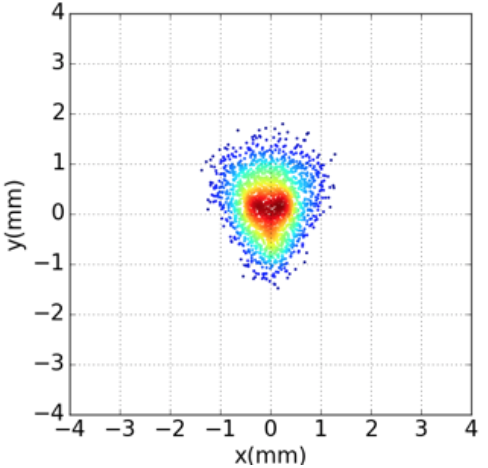}\\
			(b)
		\end{center}
	\end{minipage}
	\caption{%
		\label{fig:gun_beam_spots}%
		Beam spot at the entrance to the accelerating structure at a voltage at the control electrode and beam current:
		(a): +13~kV (330~mA),
		(b): -14~kV (50~mA).
	}
\end{figure}

\begin{figure}[thbp]
	\begin{minipage}[t]{0.49\linewidth}
		\begin{center}
			\includegraphics[width=1\linewidth]{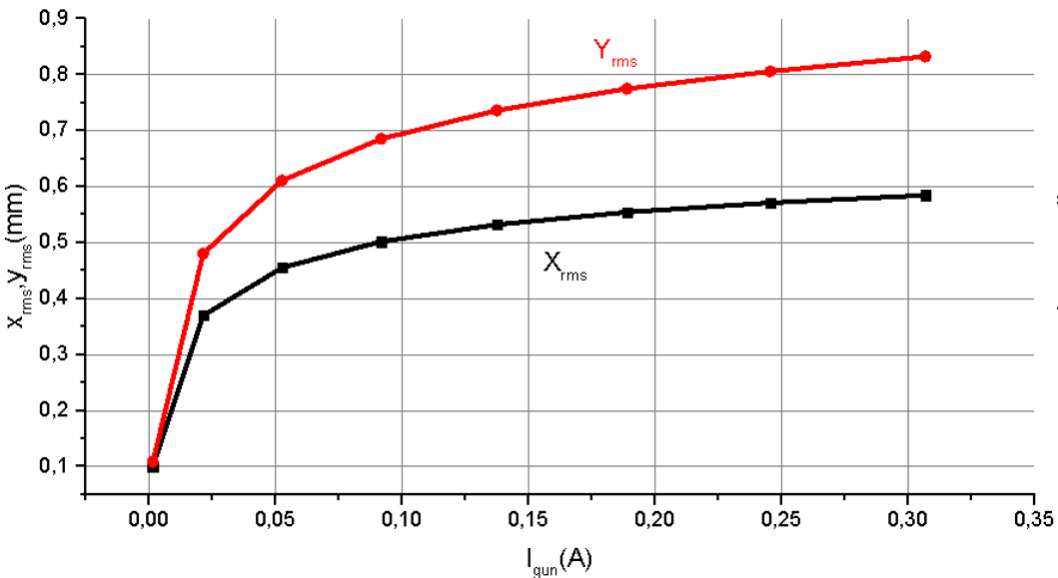}\\
			(a)
		\end{center}
	\end{minipage}
	\hfill
	\begin{minipage}[t]{0.49\linewidth}
		\begin{center}
			\includegraphics[width=1\linewidth]{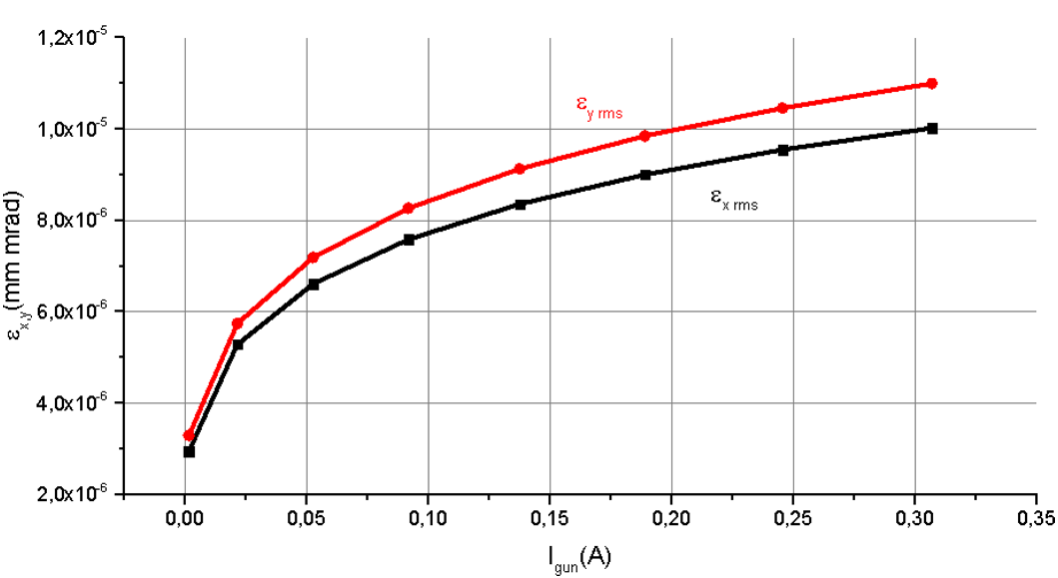}\\
			(b)
		\end{center}
	\end{minipage}
	\vspace{1pt}
	\begin{minipage}[t]{0.49\linewidth}
		\begin{center}
			\includegraphics[width=1\linewidth]{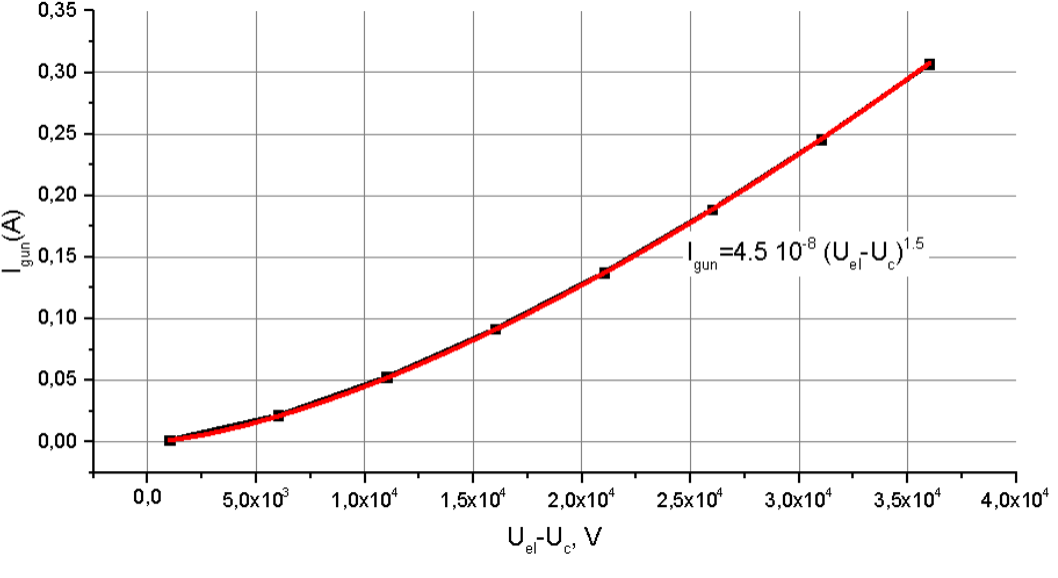}\\
			(c)
		\end{center}
	\end{minipage}
	\caption{%
		\label{fig:gun_current_dependences}%
		Dependence on the beam current
		(a) --- the rms size of the beam at the entrance to the accelerating structure,
		(b) --- normalized emittances.
		(c) Current-voltage characteristic of the gun.
	}
\end{figure}

Particle trajectories for gun currents providing the operating dose rate of bremsstrahlung at extreme energies of 6~MeV and 20~MeV are shown in Fig.~\ref{fig:gun_trajectories}, beam spots --- in Fig.~\ref{fig:gun_beam_spots}.
Figure~\ref{fig:gun_current_dependences} shows the dependencies on the current of the rms beam size and of normalized emittances.
In Fig.~\ref{fig:gun_current_dependences}~(c) shows the current-voltage characteristic of the gun.

\section{Accelerating structure}

In order to increase the acceleration efficiency and reduce the dimensions, we use standing wave C-band accelerating structure with on-axis coupling cells, operating at $\pi / 2$ mode.
The regular part of the accelerating structure is identical to the regular part of the accelerating structure of projects
\cite{Ovchinnikova__2018__Calculation_of_Electron-Beam_Dynamics_in_a_C_Band_Accelerator_for_a_Radiotherapy_Complex, Kubyshin__2010__Current_Status_of_the_12_MeV_UPC_Race_Track_Microtron}.
The main parameters of a regular accelerating cell ($\beta = 1$) are shown in Table~\ref{tab:regular_cell_parameters}.

\begin{table}[thbp]%
	\caption{\label{tab:regular_cell_parameters}%
		Parameters of a regular accelerating cell \\ with $\beta = 1$.
	}
	\begin{ruledtabular}
		\begin{tabular}{l c}
			\textrm{Parameter}&
			\textrm{Value}\\
			\colrule
			$\pi / 2$ mode frequency & 5712~MHz\\     
			Effective shunt impedance with coupling slots & 98~MOhm/m\\    
			Quality factor & 7100\\     
			Coupling factor & 10.6~\%\\      
			Beam channel diameter & 8~mm\\      
			Number of regular cells & 15\\     
			Average on-axis electric field & 30.3~MV/m\\      
			Over strength factor (max field & \\
			at surface/average field on axis) & 4.2\\   
			RF losses in accelerating cell for nominal field & 157~kW\\
		\end{tabular}
	\end{ruledtabular}
\end{table}

The irregular part of the accelerating structure includes three accelerating cells, providing bunching, focusing, and preliminary acceleration of the beam. The initial optimization of the parameters of the irregular part of the structure was carried out in calculations of the beam dynamics using the CST code in the two-dimensional (excluding coupling slots) approximation. The number of accelerating cells, the voltage at the accelerating gaps, the distance between the centers of the gaps, and the geometry of the cells were determined. For the found optimal distribution of the accelerating field, the coupling slots parameters reproducing it were found.

\begin{figure}[thbp]
	\begin{minipage}[t]{0.59\linewidth}
		\begin{center}
			\includegraphics[width=1\linewidth]{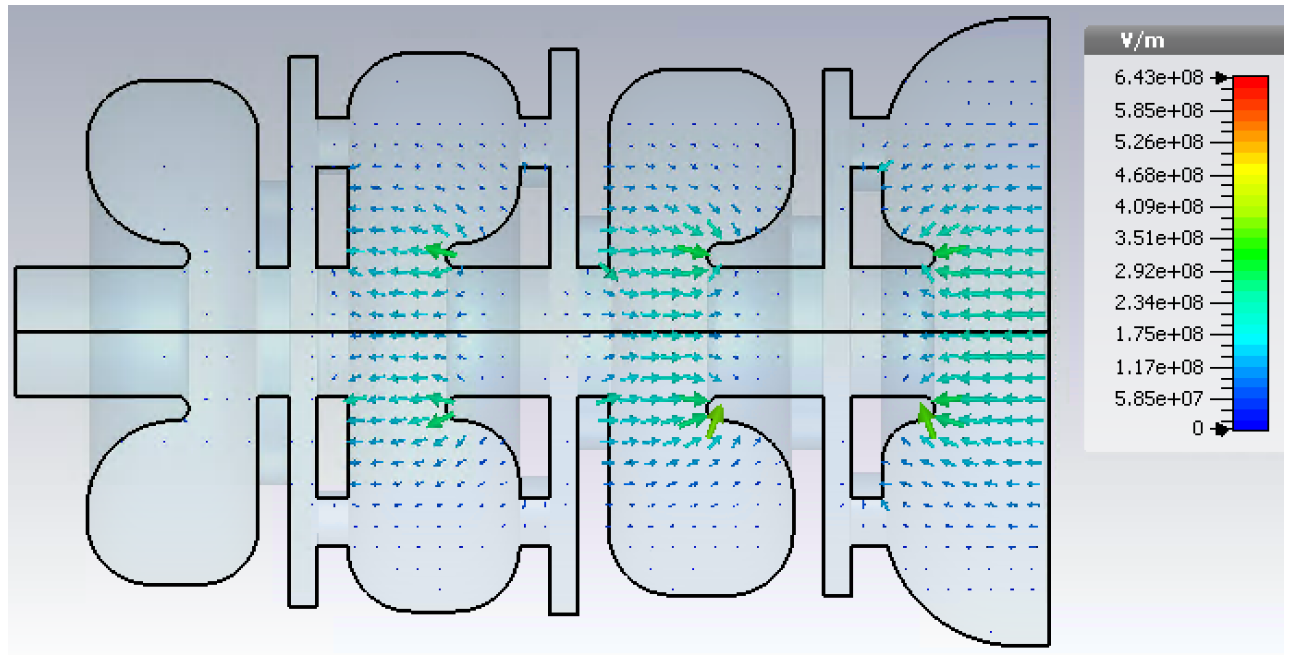}\\
			(a)
		\end{center}
	\end{minipage}
	\hfill
	\begin{minipage}[t]{0.39\linewidth}
		\begin{center}
			\includegraphics[width=1\linewidth]{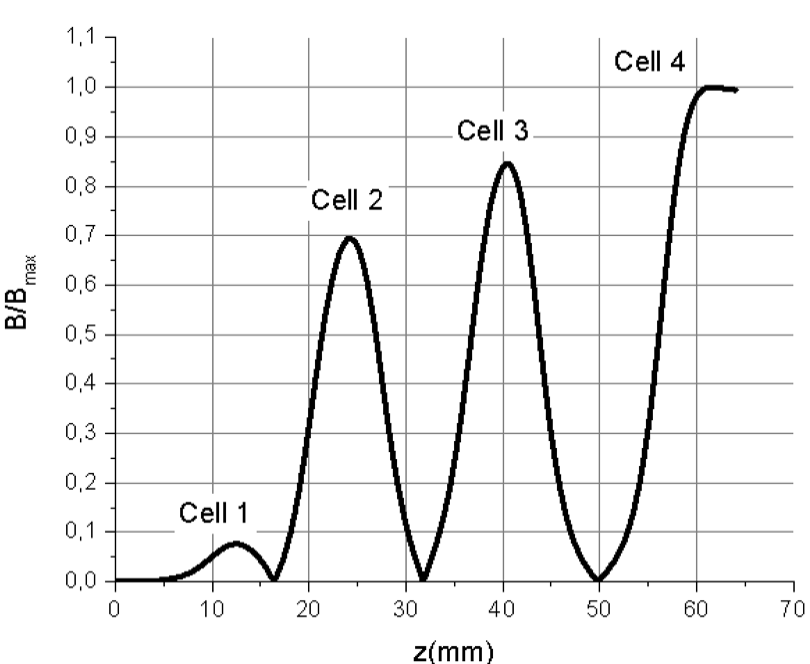}\\
			(b)
		\end{center}
	\end{minipage}
	\caption{%
		\label{fig:acc_initial_part}%
		The distribution of the electric field in the volume
		(a) and on the axis
		(b) of the irregular part of the accelerating structure.
	}
\end{figure}

The whole accelerating structure obtained as a result of optimization contains 15 cells with $\beta = 1$, including the end cell, and 3 cells with $\beta<1$, has an electric length of 440~mm.
Fig.~\ref{fig:acc_initial_part} shows the geometry of the cells of the initial part and the distribution of the accelerating field on the axis.

The total RF power loss in the walls to get an energy of $E_b = 10 \text{~MeV}$ is $P_w \approx 2.66 \text{~MW}$, the effective shunt impedance of the entire structure is $Z_{eff} = E_b^2/P_w \approx 37.6 \text{~M}\Omega$.
The maximum electric field strength at the surface of the structure is about 130~MV/m.

\section{Magnetic mirror}

\subsection{Beam optics}

\begin{figure}[thbp]
	\begin{center}
		\includegraphics[width=1\linewidth]{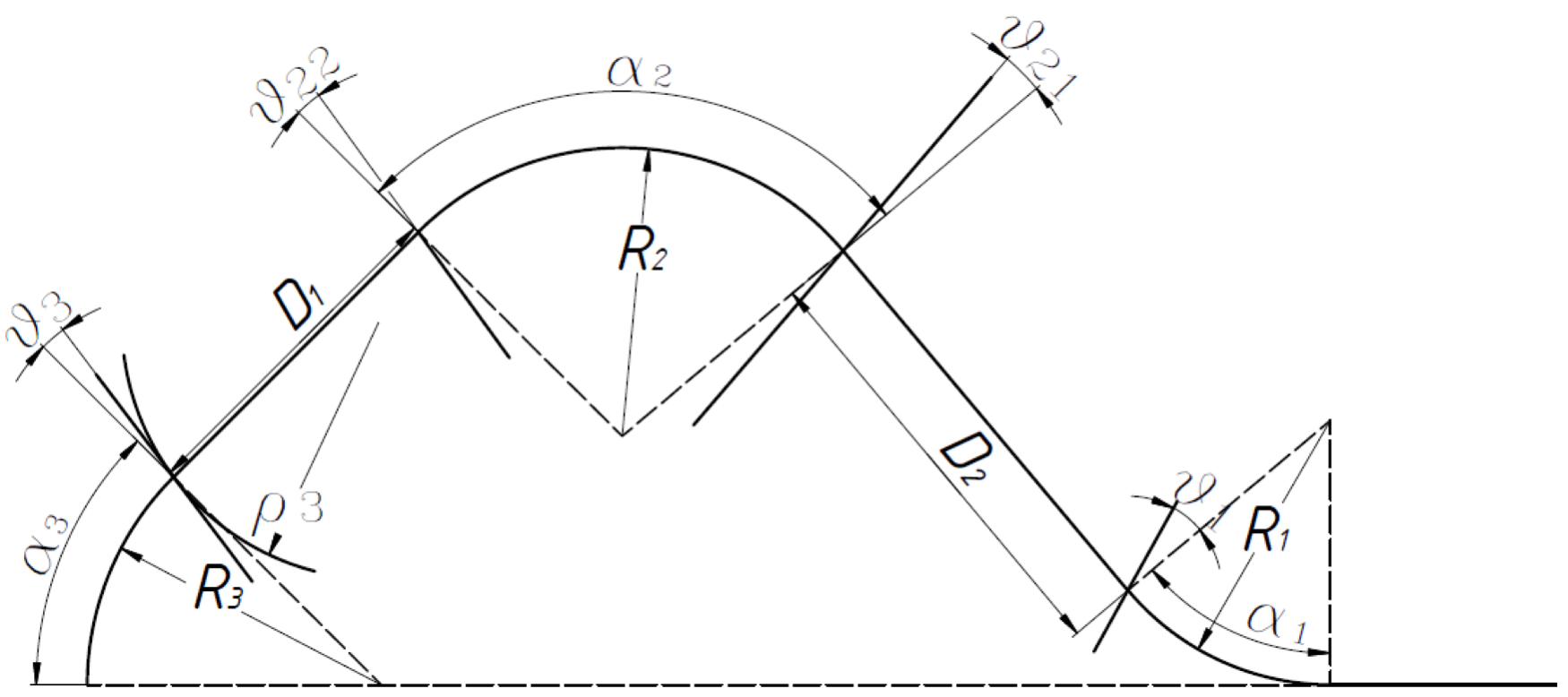}
		\caption{%
			\label{fig:magnetic_mirror_parameterization}%
			Parameterization of the magnetic mirror.
		}
	\end{center}
\end{figure}

Five-magnet mirror is shown in Fig.~\ref{fig:accelerator_scheme}.
The system is mirror-symmetric with respect to a plane passing through the axis of the accelerating structure, with input and output magnets structurally combined \cite{Schriber__1975__Double_Pass_Linear_Accelerator_Reflexotron, Schriber__1977__Experimental_Measurements_on_a_25_Mev_Reflexotron, Taylor__1983__Therac_25_A_New_Medical_Accelerator_Concept}.
The parameterization of the magnetic mirror is shown in Fig.~\ref{fig:magnetic_mirror_parameterization}.

The parameters of the magnetic mirror are related by (\ref{eq:magnetic_mirror_parameterization__a3}) and (\ref{eq:magnetic_mirror_parameterization__d1}):
\begin{equation}
\alpha_3 = \pi / 2 - \alpha_2 + \alpha_1
\label{eq:magnetic_mirror_parameterization__a3}
\end{equation}
\begin{widetext}
\begin{equation}
d_1=\frac{d_2\sin(\alpha_1)+(R_1+R_2)(1-\cos(\alpha_1))-R_2+(R_2-R_3)\cos(\alpha_2-\alpha_1)}{\sin(\alpha_2-\alpha_1)}
\label{eq:magnetic_mirror_parameterization__d1}
\end{equation}
\end{widetext}
where
$\alpha_1, \alpha_2, \alpha_3$ are the bending angles,
$R_1, R_2, R_3$ are the radii of curvature of central trajectory, 
$d_1, d_2$ are the distances between the effective boundaries of the magnetic field.
The angles of pole face rotation are designated as
$\vartheta_1, \vartheta_{21}, \vartheta_{22}, \vartheta_3$.

Optics of the magnetic mirror was optimized using the MAD-X code \cite{MAD_X} for momentum of central trajectory of 10.5~MeV/c.
In the first order, the following conditions were imposed on the elements of the transfer matrix:
$R_{21}=0, R_{43}=0, R_{16}=0$ and $R_{26}=0$, which corresponds to an infinite focal length in the vertical and horizontal planes, as well as zero spatial and angular dispersion.
The matrix element $R_{16}$ was optimized to provide maximum bunch compression in the longitudinal phase space.

Bending radii were chosen to provide small size of the system at a reasonably high magnetic field induction in the gaps, taking into account relations (\ref{eq:magnetic_mirror_parameterization__a3}) and (\ref{eq:magnetic_mirror_parameterization__d1}).
Variable parameters were
$d_1, \vartheta_1, \vartheta_{21}, \vartheta_{22}, \vartheta_3$.

The optimization of the magnetic mirror was carried out iteratively in several steps, considering the results of magnetic system and beam dynamics calculations.
In particular, the value of the fringe field coefficients and the effective lengths were refined by the magnets calculation results; the second-order optimization was performed according to the results of beam dynamics calculation, and the radii of the pole face curvature of the third magnet were introduced.
The final parameters of the mirror magnets are given in Table~\ref{tab:mirror_final_parameters}.

\begin{table}[thbp]%
	\caption{\label{tab:mirror_final_parameters}%
		Final parameters of mirror magnets.
	}
		\begin{tabular}{l c}
			\hline
			\hline
			\textrm{Parameter}&
			\textrm{Value}\\
			\hline
			$\alpha_1$ & $50^{\circ}$ \\
			$\alpha_2$ & $95^{\circ}$ \\    
			$\alpha_3$ & $45^{\circ}$ \\     
			$R_1$ & 42.16~mm \\     
			$R_2$ & 45.955~mm\\    
			$R_3$ & 46.975~mm\\     
			$\vartheta_1$ & $21.1^{\circ}$ \\     
			$\vartheta_{21}$ & $9.4^{\circ}$ \\     
			$\vartheta_{22}$ & $-9.0^{\circ}$ \\ 
			$\vartheta_3$ & $-7.73^{\circ}$ \\     
			$\rho_3$ & 98~mm \\ 
			\hline
			\hline
		\end{tabular}
\end{table}

\subsection{Magnets calculation}

In order to unambiguously control the beam energy during the first acceleration, the magnetic mirror is built on the basis of rare-earth permanent magnet material.
With a constant magnetic field, a deviation of the beam energy from the design value of 10~MeV will lead to an increase in the loss of the beam in the mirror and the triggering of an interlock switching off the beam.

The Halbach box-type design \cite{Novikov__1998__Novel_race_track_microtron_end_magnets}, previously used by us in the development of the race-track microtron magnets \cite{Shvedunov__2005__A_70_Mev_racetrack_microtron, Kubyshin__2010__Current_Status_of_the_12_MeV_UPC_Race_Track_Microtron}, was taken as the basis for the design of mirror magnets.
The essence of the design is the concentration of magnetic flux into the pole gap from the blocks of rare-earth permanent magnets located at all sides of the pole, except for the side of the gap, and closed by the yoke of the box structure.

The design of the magnetic mirror with a common yoke for all magnets is shown in Fig.~\ref{fig:accelerator_scheme}.
The magnitude of the magnetic field induction in the pole gap with a height of 20~mm is 0.831~T, 0.762~T and 0.746~T, respectively, for magnets M1, M2 and M3.

\begin{figure}[thbp]
	\begin{minipage}[t]{0.39\linewidth}
		\begin{center}
			\includegraphics[width=1\linewidth]{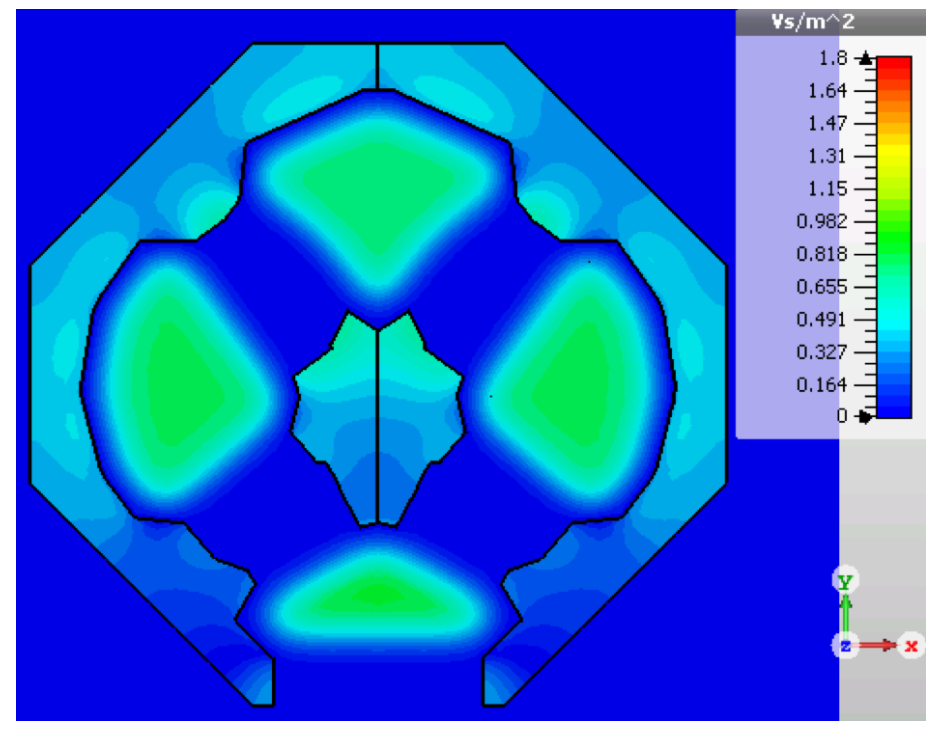}\\
			(a)
		\end{center}
	\end{minipage}
	\hfill
	\begin{minipage}[t]{0.59\linewidth}
		\begin{center}
			\includegraphics[width=1\linewidth]{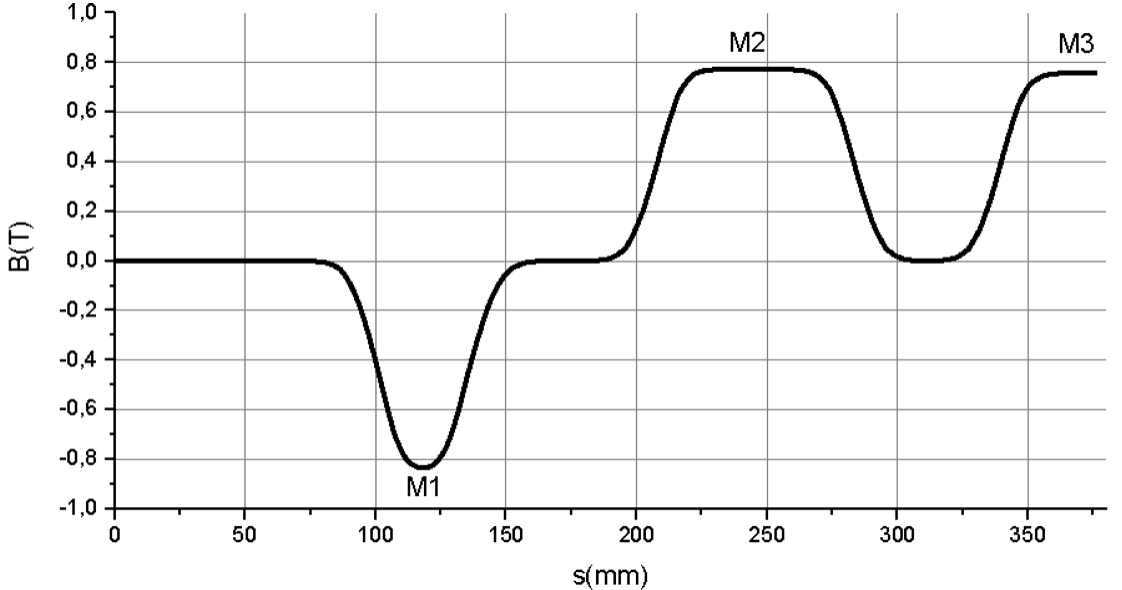}\\
			(b)
		\end{center}
	\end{minipage}
	\caption{%
		\label{fig:mirror_magnetic_field}%
		(a) Two-dimensional distribution of the magnetic field induction in the median plane of the magnetic mirror, calculated by CST code;
		(b) the distribution of the magnetic field induction along the first half of the central trajectory.
	}
\end{figure}

The initial approximation for the rare-earth permanent magnet material blocks dimension was chosen in accordance with \cite{Novikov__1998__Novel_race_track_microtron_end_magnets}, while the residual magnetization of all blocks was taken equal to $B_r = 1 \text{~T}$.
The calculations were performed using the CST code.
The calculated two-dimensional distribution of the magnetic field induction in the median plane of the magnetic mirror is shown in Fig.~\ref{fig:mirror_magnetic_field}~(a).
The dimension of the poles was adjusted to ensure the equality of the effective lengths of the magnets obtained in CST and in the MAD-X codes.
Fig.~\ref{fig:mirror_magnetic_field}~(b) shows the final distribution of the magnetic field induction in the median plane along the first half of the central trajectory, providing the optical properties of the mirror close to the properties obtained in MAD-X.
To reduce aberrations caused by second-order effects, a rounding of the pole face was introduced in the third magnet.

In this paper, we do not consider design features, techniques of manufacturing, assembly, and tuning of magnets.
Obviously, in order to accurately set the values of the magnetic field induction in the median plane during the tuning process, individual magnets must contain tuning elements of one or another design.
These issues, in particular, were considered in \cite{Vladimirov__2014__End_magnets_with_rare_earth_permanent_magnet_material_for_a_compact_race-track_microtron}.

\section{270-degree bending magnet}

A $270^{\circ}$ magnet bends an electron beam with energy varying in the range of 6 -- 20~MeV, and therefore cannot be built on the basis of permanent magnets.
Of the various options described in \cite{Karzmark__1993__Medical_Electron_Accelerators}~(p.~116), we selected the classical three-magnet system \cite{Brown__1975__Achromatic_magnetic_beam_deflection_system}, shown in Fig.~\ref{fig:accelerator_scheme}, in which each of the magnets bends the beam by $90^{\circ}$, while the system is mirror symmetric with respect to the plane passing through the center of the second magnet.

The optics of the magnetic system was optimized using the MAD-X code for radii of curvature of central trajectory of 65~mm, the same for all magnets, which corresponds to magnetic field induction in the pole gap of 1.05~T for an energy of 20~MeV.
The height of the pole gap is taken rather large in comparison with the known analogues, equal to 20~mm.

One of the basic requirements for a $270^{\circ}$ magnet is zero spatial and angular dispersion at the bremsstrahlung target, $R_{16} = R_{26} = 0$.
In addition, the asymmetry of the beam spot resulting from the absence of axial symmetry of the electron gun and the magnetic mirror was taken into account during the optimization.
Minimization of the beam asymmetry at the target was ensured by the choice of the values of the transfer matrix elements $R_{21}$ and $R_{43}$, which determine the focusing properties of the system. 
Finally, in order to efficiently select particles by energy using collimator \textit{6} (Fig.~\ref{fig:accelerator_scheme}), the spatial dispersion in the symmetry plane of the system was chosen maximum, and here the beam crossover was located in the bending plane.
The angles of pole face rotation obtained as a result of optimization are given in Table~\ref{tab:bending_angles}.
The drift space between the effective boundaries of the magnets M1 -- M2, as well as M2 -- M3 is 55~mm.

\begin{table}[thbp]%
	\caption{\label{tab:bending_angles}%
		Values of pole face rotation angles.
	}
	\begin{ruledtabular}
		\begin{tabular}{l c c}
			Magnet &
			Entrance pole &
			Exit pole \\
			 &
			face rotation angle &
			face rotation angle \\
			\colrule
			M1 & $-2^{\circ}$  & $37.2^{\circ}$\\      
			M2 & $-6.15^{\circ}$  & $-6.15^{\circ}$\\     
			M3 & $37.2^{\circ}$ & $-2^{\circ}$\\
		\end{tabular}
	\end{ruledtabular}
\end{table}

Calculation of the magnetic field for the geometry of the system shown in Fig.~\ref{fig:accelerator_scheme}, was performed with CST code.
In contrast to similar systems described by other authors, in order to reduce the degree of overlap of the fringe fields of neighboring magnets with a large pole gap height chosen by us, each of the magnets has individual windings.

\begin{figure}[thbp]
	\begin{minipage}[t]{0.425\linewidth}
		\begin{center}
			\includegraphics[width=1\linewidth]{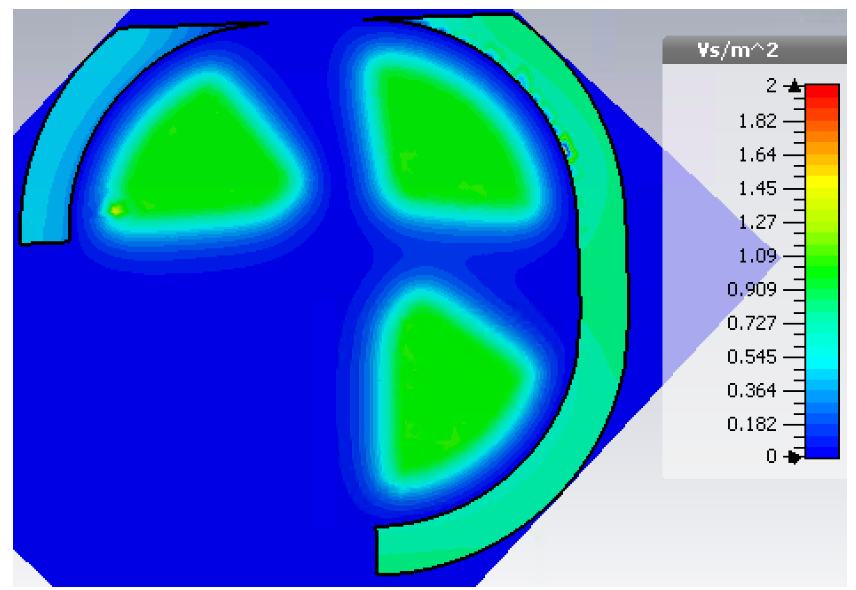}\\
			(a)
		\end{center}
	\end{minipage}
	\hfill
	\begin{minipage}[t]{0.555\linewidth}
		\begin{center}
			\includegraphics[width=1\linewidth]{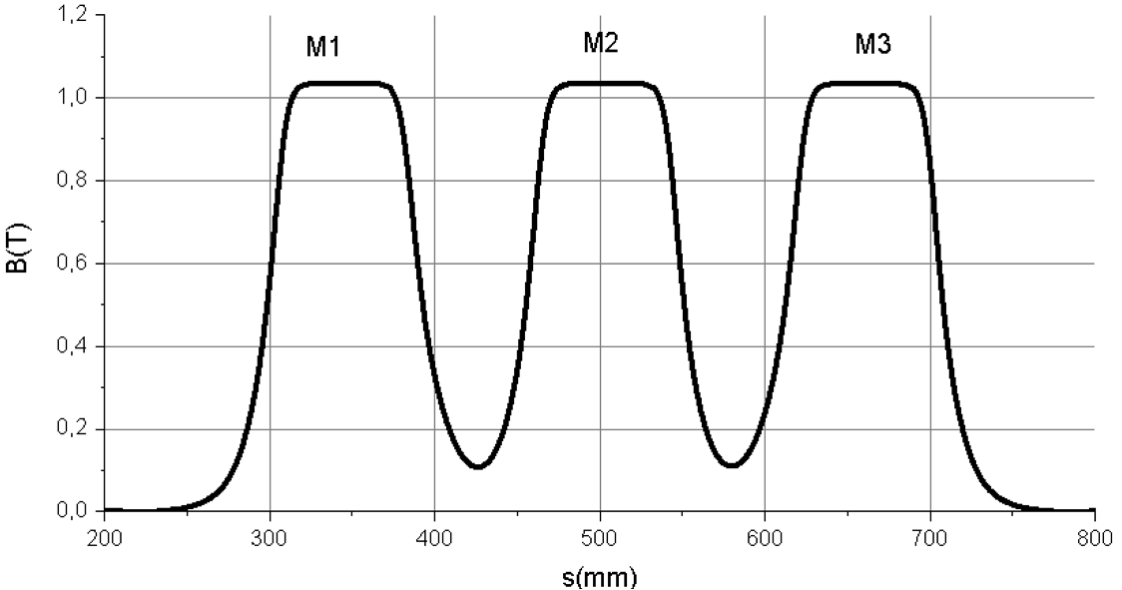}\\
			(b)
		\end{center}
	\end{minipage}
	\caption{%
		\label{fig:bending_magnetic_field}%
		Distribution of the magnetic field induction in the median plane, calculated by CST code:
		(a) --- 2D,
		(b) --- along the central trajectory.
	}
\end{figure}

In Fig.~\ref{fig:bending_magnetic_field}~(a) the two-dimensional distribution of the magnetic field induction in the median plane is shown, and in Fig.~\ref{fig:bending_magnetic_field}~(b) along the central trajectory.
It is seen that the fringe fields of the individual magnets noticeably overlap.
To ensure the properties of the system close to those obtained during optimization with MAD-X code, the dimensions of the poles were adjusted.

\section{Beam dynamics}

Beam dynamics simulation with CST code was carried out sequentially at all stages of accelerator optimization.
Following the results of simulation, the parameters of the electron gun, accelerating structure, magnetic mirror and a $270^{\circ}$ magnet were adjusted, after which the beam dynamics simulation was repeated. 
This section presents the final results.

\subsection{Bremsstrahlung mode}

As shown in Section~\ref{sec:Accelerator_System_Requirements}, to provide project dose rate to each value of the final energy must correspond definite gun current, varying from about 330~mA for energy of 6~MeV to about 50~mA for 20~MeV.
Simulation of the beam dynamics were performed over the entire energy range with voltage at the gun’s control electrode, providing for each energy the required accelerated beam current.

\begin{figure*}[t!]
	\begin{minipage}[t]{0.23\linewidth}
		\begin{center}
			\includegraphics[width=1\linewidth]{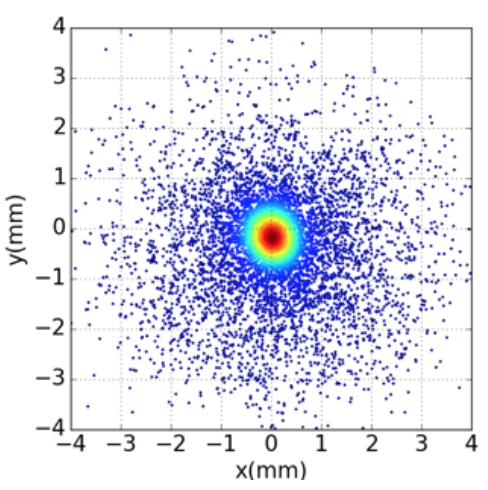}\\
			(a)
		\end{center}
	\end{minipage}
	\hfill
	\begin{minipage}[t]{0.23\linewidth}
		\begin{center}
			\includegraphics[width=1\linewidth]{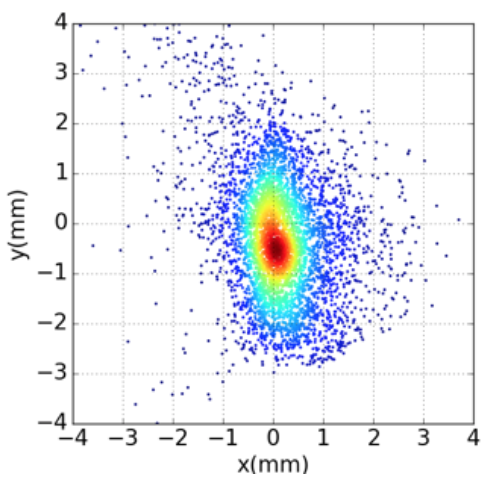}\\
			(b)
		\end{center}
	\end{minipage}
	\hfill
	\begin{minipage}[t]{0.23\linewidth}
		\begin{center}
			\includegraphics[width=1\linewidth]{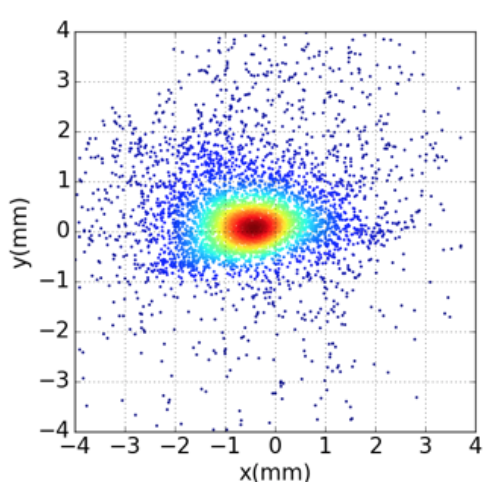}\\
			(c)
		\end{center}
	\end{minipage}
	\hfill
	\begin{minipage}[t]{0.26\linewidth}
		\begin{center}
			\includegraphics[width=1\linewidth]{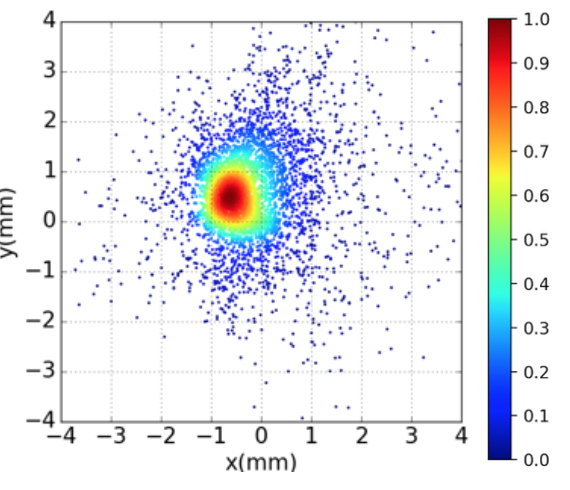}\\
			(d)
		\end{center}
	\end{minipage}
	\caption{%
		\label{fig:bremsstrahlung_mode__beam_spots__6}%
		The beam spots in bremsstrahlung mode for a final energy of 6~MeV at an injection current of 330~mA:
		(a) --- at the output of the accelerating structure after the first passage;
		(b) --- at the output of the magnetic mirror;
		(c) --- at the exit of the accelerating structure after second passage;
		(d) --- at the bremsstrahlung target.
	}
\end{figure*}

\begin{figure*}[t!]
	\begin{minipage}[t]{0.23\linewidth}
		\begin{center}
			\includegraphics[width=1\linewidth]{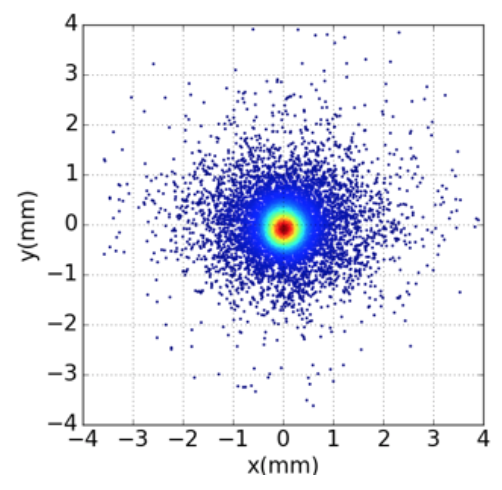}\\
			(a)
		\end{center}
	\end{minipage}
	\hfill
	\begin{minipage}[t]{0.23\linewidth}
		\begin{center}
			\includegraphics[width=1\linewidth]{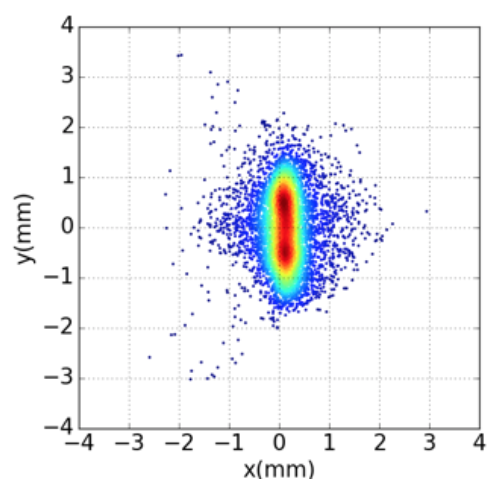}\\
			(b)
		\end{center}
	\end{minipage}
	\hfill
	\begin{minipage}[t]{0.23\linewidth}
		\begin{center}
			\includegraphics[width=1\linewidth]{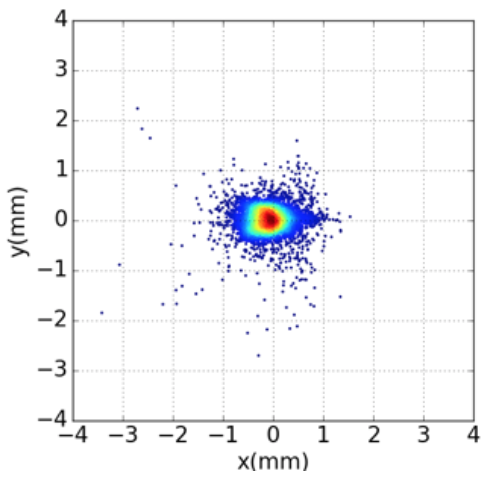}\\
			(c)
		\end{center}
	\end{minipage}
	\hfill
	\begin{minipage}[t]{0.26\linewidth}
		\begin{center}
			\includegraphics[width=1\linewidth]{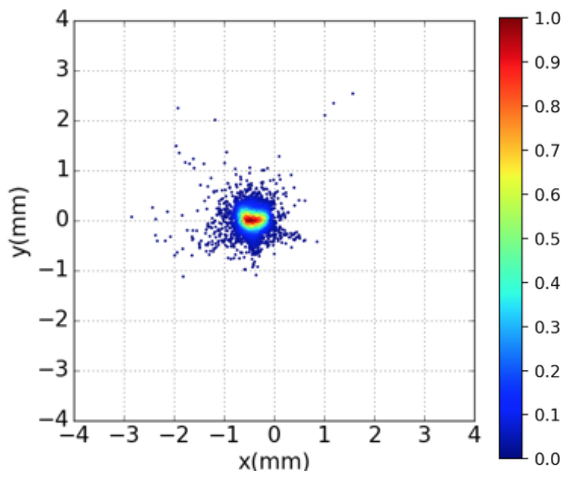}\\
			(d)
		\end{center}
	\end{minipage}
	\caption{%
		\label{fig:bremsstrahlung_mode__beam_spots__20}%
		The beam spots in bremsstrahlung mode for a final energy of 20 MeV at an injection current of 50~mA:
		(a) --- at the output of the accelerating structure after the first passage;
		(b) --- at the output of the magnetic mirror;
		(c) --- at the exit of the accelerating structure after second passage;
		(d) --- at the bremsstrahlung target.
	}
\end{figure*}

In Fig.~\ref{fig:bremsstrahlung_mode__beam_spots__6}, the beam spots at the output of the accelerating structure after its first passage, at the output of the magnetic mirror, at the output of the accelerating structure after second passage and at the bremsstrahlung target are shown for the accelerated beam energy of 6~MeV and the injection current of 330~mA.
Similar data are shown in Fig.~\ref{fig:bremsstrahlung_mode__beam_spots__20} for an energy of 20~MeV and an injection current of 50~mA.

\begin{figure*}[t!]
	\begin{minipage}[t]{0.23\linewidth}
		\begin{center}
			\includegraphics[width=1\linewidth]{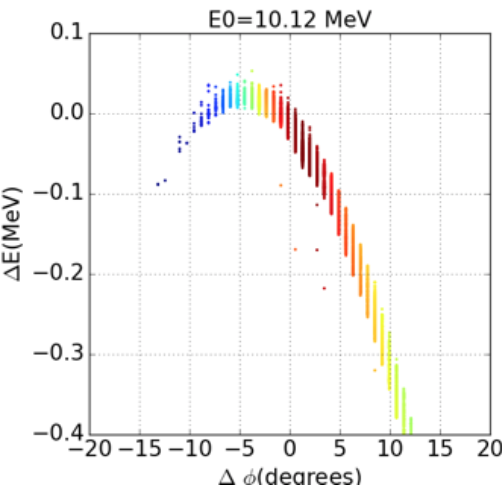}\\
			(a)
		\end{center}
	\end{minipage}
	\hfill
	\begin{minipage}[t]{0.23\linewidth}
		\begin{center}
			\includegraphics[width=1\linewidth]{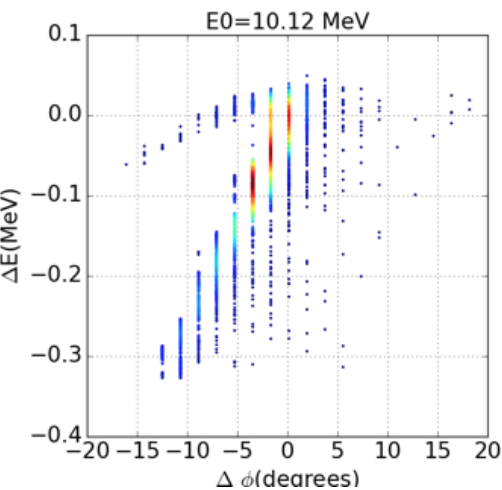}\\
			(b)
		\end{center}
	\end{minipage}
	\hfill
	\begin{minipage}[t]{0.23\linewidth}
		\begin{center}
			\includegraphics[width=1\linewidth]{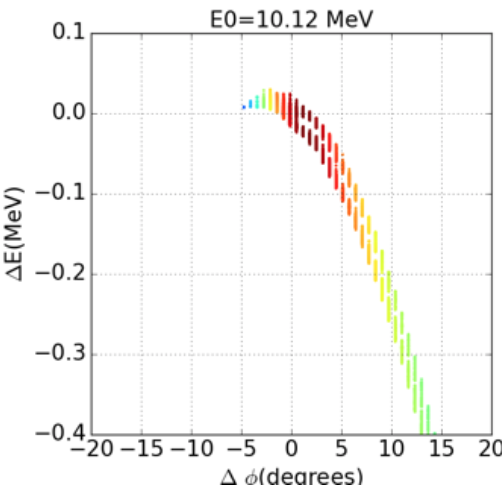}\\
			(c)
		\end{center}
	\end{minipage}
	\hfill
	\begin{minipage}[t]{0.26\linewidth}
		\begin{center}
			\includegraphics[width=1\linewidth]{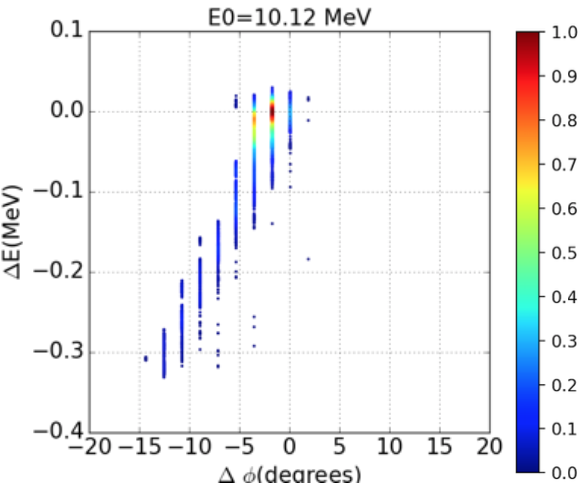}\\
			(d)
		\end{center}
	\end{minipage}
	\caption{%
		\label{fig:bremss_mode__long_phase_spaces}%
		Longitudinal phase spaces in the bremsstrahlung mode:
		(a), (b) --- for a final energy of 6~MeV at an injection current of 330~mA at the output of the accelerating structure after the first passage and at the output of the magnetic mirror;
		(c), (d) --- similar data for a final energy of 20~MeV and the injection current of 50~mA.
	}
\end{figure*}

As discussed in Section~\ref{sec:Accelerator_scheme}, the width of the energy spectrum of an accelerated beam varies with energy and depends on the phase length of the bunch during its second acceleration.
In Fig.~\ref{fig:bremss_mode__long_phase_spaces} longitudinal phase spaces are shown at the injection current of 330~mA (for 6~MeV) at the output of the accelerating structure after the first passage and at the output of a magnetic mirror, and similar data are shown for injection current of 50~mA (for 20~MeV).
As follows from Fig.~\ref{fig:bremss_mode__long_phase_spaces}~(a, c), the longitudinal phase space at the input of the magnetic mirror is essentially nonlinear due to the sinusoidal time dependence of the accelerating field, and the degree of nonlinearity depends on the injection current.


\begin{figure*}[t!]
	\begin{minipage}[t]{0.24\linewidth}
		\begin{center}
			\includegraphics[width=1\linewidth]{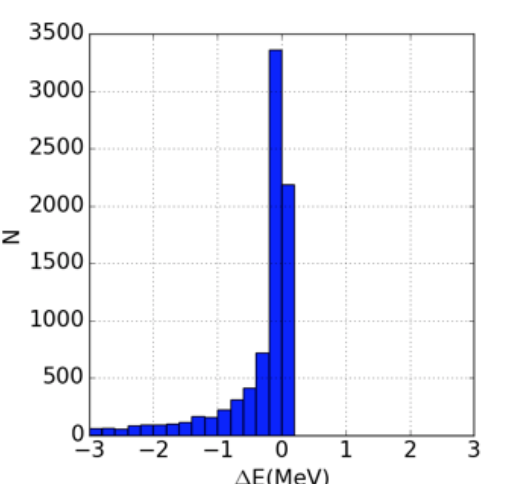}\\
			(a)
		\end{center}
	\end{minipage}
	\hfill
	\begin{minipage}[t]{0.24\linewidth}
		\begin{center}
			\includegraphics[width=1\linewidth]{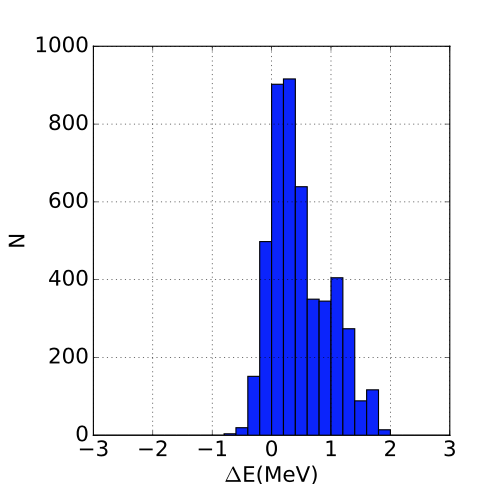}\\
			(b)
		\end{center}
	\end{minipage}
	\hfill
	\begin{minipage}[t]{0.24\linewidth}
		\begin{center}
			\includegraphics[width=1\linewidth]{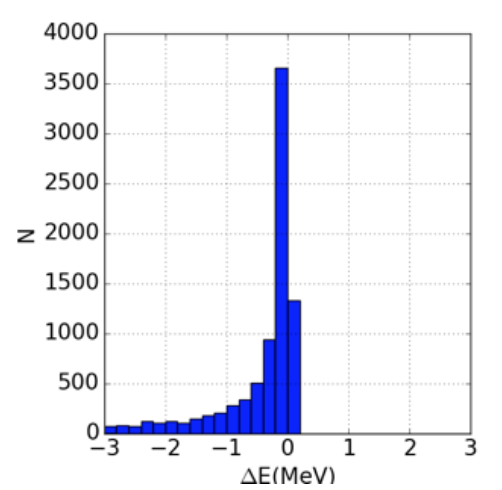}\\
			(c)
		\end{center}
	\end{minipage}
	\hfill
	\begin{minipage}[t]{0.24\linewidth}
		\begin{center}
			\includegraphics[width=1\linewidth]{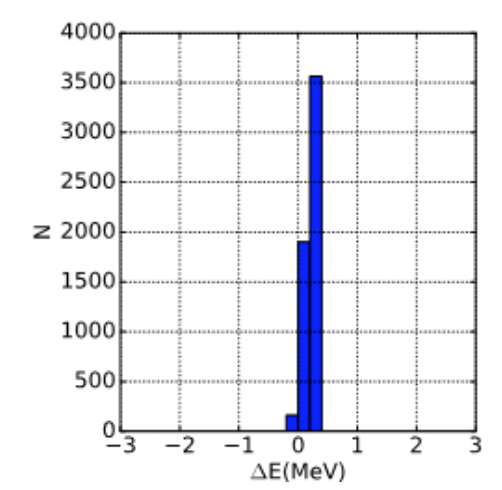}\\
			(d)
		\end{center}
	\end{minipage}
	\caption{%
		\label{fig:bremss_mode__spectra}%
		Beam spectra in the bremsstrahlung mode:
		(a), (b) --- for a final energy of 6~MeV at an injection current of 330~mA at the output of the accelerating structure after the first and second passage;
		(c), (d) --- similar data for a final energy of 20~MeV and the injection current of 50~mA.
	}
\end{figure*}

\begin{figure}[t!]
	\begin{minipage}[t]{0.49\linewidth}
		\begin{center}
			\includegraphics[width=1\linewidth]{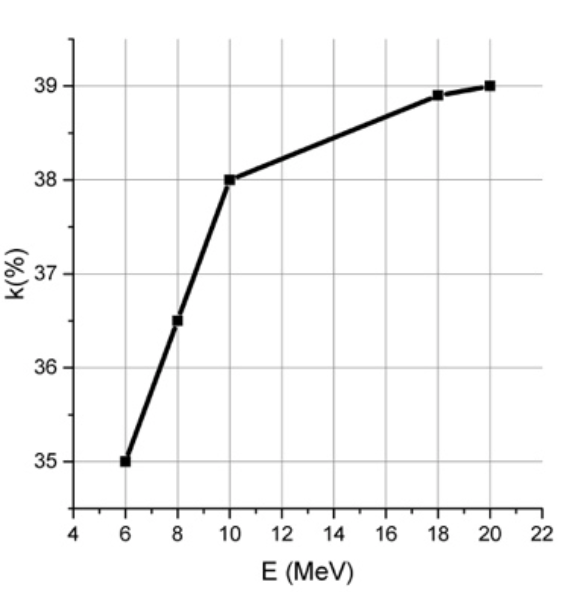}\\
			(a)
		\end{center}
	\end{minipage}
	\hfill
	\begin{minipage}[t]{0.49\linewidth}
		\begin{center}
			\includegraphics[width=1\linewidth]{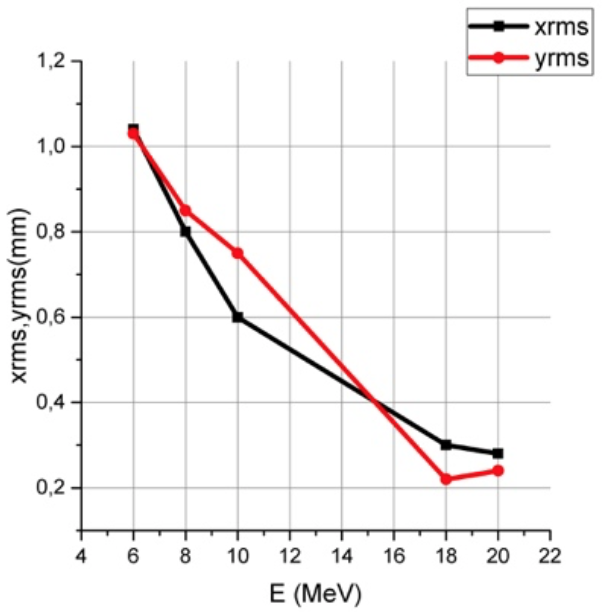}\\
			(b)
		\end{center}
	\end{minipage}
	\vspace{1pt}
	\begin{minipage}[t]{0.49\linewidth}
		\begin{center}
			\includegraphics[width=1\linewidth]{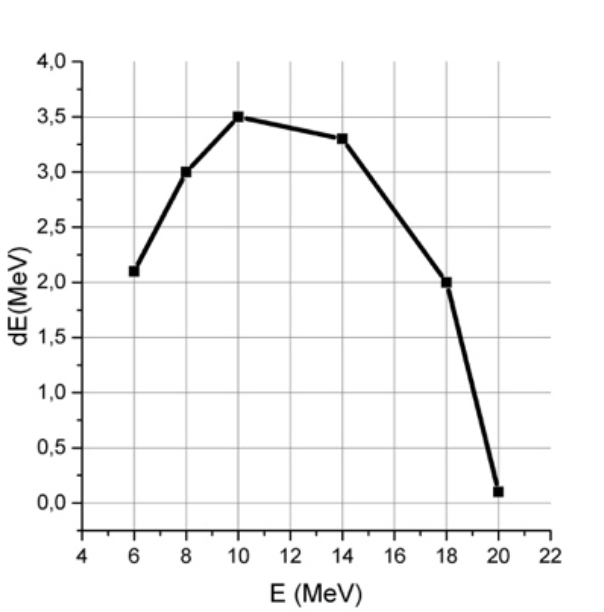}\\
			(c)
		\end{center}
	\end{minipage}
	\caption{%
		\label{fig:bremss_target__dependence}%
		Dependence on the energy of the accelerated beam at the bremsstrahlung target:
		(a) the fraction of injected particles reaching the target,
		(b) the rms size of the beam in two perpendicular planes,
		(c) the total energy spread.
	}
\end{figure}

The nonlinearity of the phase space limits the possibility to compress the bunch length using longitudinal dispersion of the magnetic mirror. 
The degree of nonlinearity could be reduced by shifting the acceleration phase of the bunches from the phase of maximum energy gain during first passage of accelerating structure, however, the acceleration efficiency would decrease and the RF power consumption would increase.
In this regard, the longitudinal dispersion of the mirror was chosen from the condition of optimal bunch compression over the whole range the injection current.
The low-energy tail was cut off by collimator \textit{5} (Fig.~\ref{fig:accelerator_scheme}) at the level of 3\% of the spectrum maximum.

The longitudinal phase spaces of the beam after the magnetic mirror for the two discussed values of the energy and injection current are shown in Fig.~\ref{fig:bremss_mode__long_phase_spaces}~(b, d).
In both cases, the total phase length of the bunches is close to $15^{\circ}$.
As can be seen from Fig.~\ref{fig:bremss_mode__spectra}, where the beam spectra are shown after the first and second passage of the accelerating structure, the total spectrum width at energy of 6~MeV exceeds 2~MeV, while for an energy of 20~MeV it is about 0.8~MeV.

Fig.~\ref{fig:bremss_target__dependence} shows the data for the entire range of energies at the position of bremsstrahlung target: the fraction of injected particles reaching the target, the rms beam sizes in two perpendicular planes, and the width of the energy spectrum.

\subsection{Electron beam mode}

\begin{figure*}[thbp]
	\begin{minipage}[t]{0.23\linewidth}
		\begin{center}
			\includegraphics[width=1\linewidth]{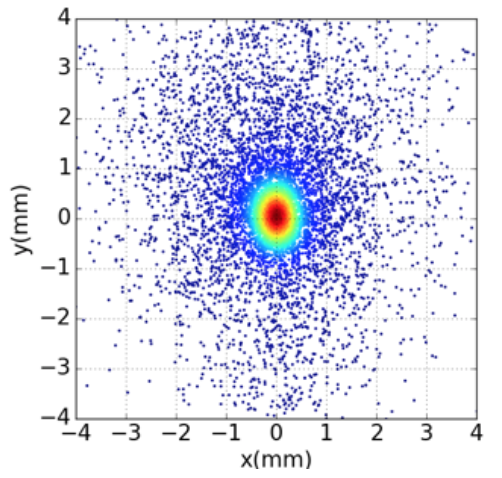}\\
			(a)
		\end{center}
	\end{minipage}
	\hfill
	\begin{minipage}[t]{0.23\linewidth}
		\begin{center}
			\includegraphics[width=1\linewidth]{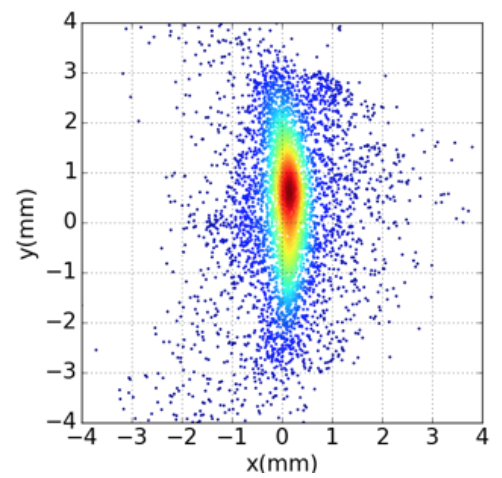}\\
			(b)
		\end{center}
	\end{minipage}
	\hfill
	\begin{minipage}[t]{0.23\linewidth}
		\begin{center}
			\includegraphics[width=1\linewidth]{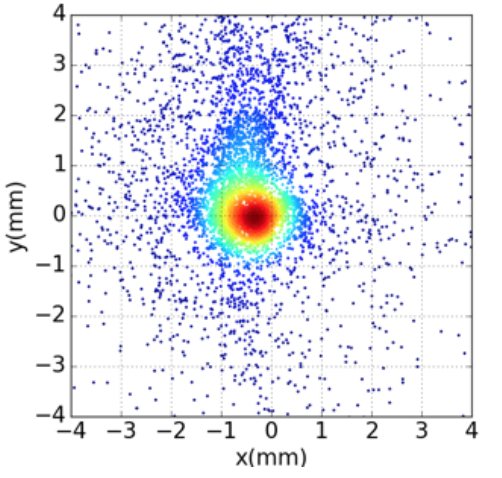}\\
			(c)
		\end{center}
	\end{minipage}
	\hfill
	\begin{minipage}[t]{0.26\linewidth}
		\begin{center}
			\includegraphics[width=1\linewidth]{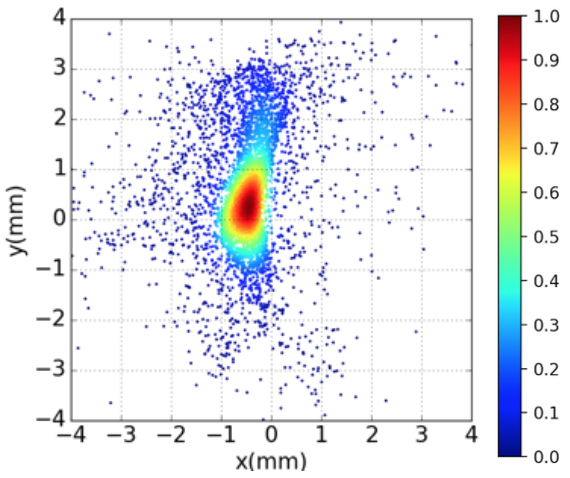}\\
			(d)
		\end{center}
	\end{minipage}
	\caption{%
		\label{fig:electron_mode__beam_spots}%
		The beam spots in the electron mode at the injection current of 0.8~mA:
		(a) --- at the output of the accelerating structure after the first passage;
		(b) --- at the output of the magnetic mirror;
		(c) --- at the exit window at the energy of 6~MeV;
		(d) --- at the exit window at the energy of 20~MeV.
	}
\end{figure*}

For operation in the electron beam mode, the accelerated current must be several orders of magnitude lower than for bremsstrahlung.
In the chosen design of the electron gun, by reducing the voltage at the control electrode relative to the cathode, the current can be decreased to about 0.8~mA while maintaining acceptable transverse beam dimensions and divergence at the entrance to the accelerating structure.
In Fig.~\ref{fig:electron_mode__beam_spots} the beam spots calculated for this value of the injection current at the output of the accelerating structure after its first passage, at the output of the magnetic mirror, and at the exit window at energies of 6 MeV and 20~MeV are shown.

\begin{figure}[thbp]
	\begin{minipage}[t]{0.45\linewidth}
		\begin{center}
			\includegraphics[width=1\linewidth]{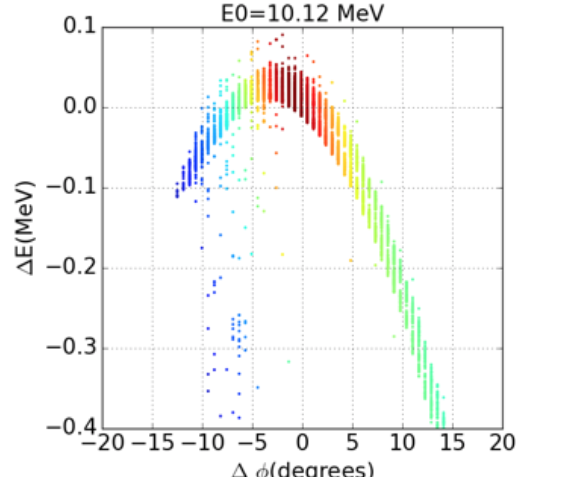}\\
			(a)
		\end{center}
	\end{minipage}
	\hfill
	\begin{minipage}[t]{0.53\linewidth}
		\begin{center}
			\includegraphics[width=1\linewidth]{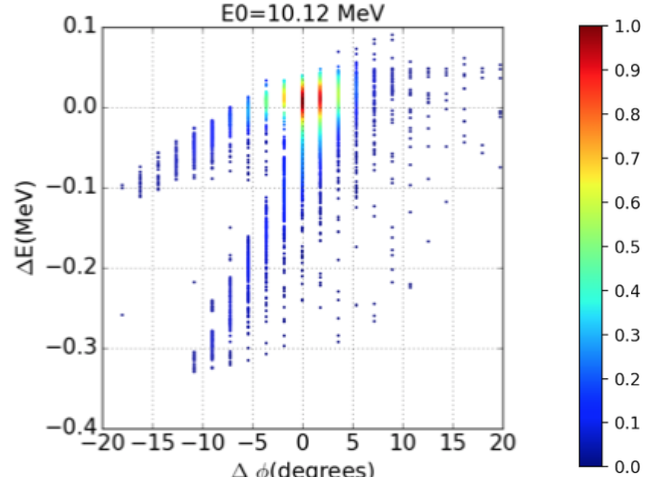}\\
			(b)
		\end{center}
	\end{minipage}
	\caption{%
		\label{fig:electron_mode__long_phase_spaces}%
		Longitudinal phase space in the electron mode at an injection current of 0.8~mA:
		(a) --- at the output of the accelerating structure after the first passage;
		(b) --- at the exit of the magnetic mirror.
	}
\end{figure}

\begin{figure}[thbp]
	\begin{minipage}[t]{0.46\linewidth}
		\begin{center}
			\includegraphics[width=1\linewidth]{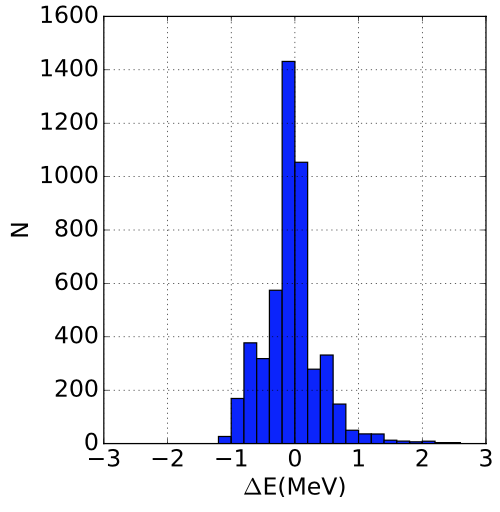}\\
			(a)
		\end{center}
	\end{minipage}
	\hfill
	\begin{minipage}[t]{0.52\linewidth}
		\begin{center}
			\includegraphics[width=1\linewidth]{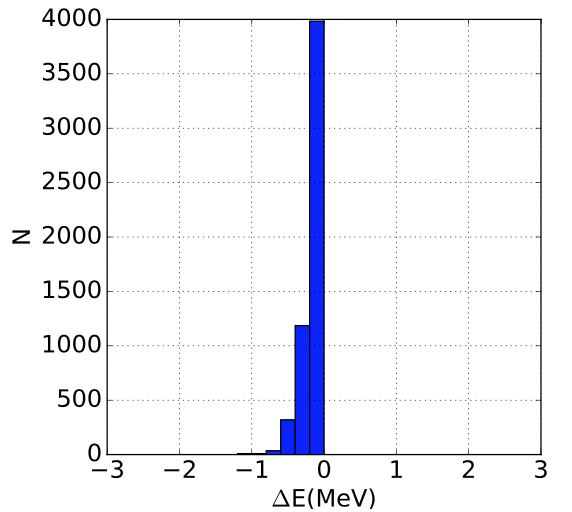}\\
			(b)
		\end{center}
	\end{minipage}
	\caption{%
		\label{fig:electron_mode__spectra}%
		Beam spectra in the electron mode at the output of the accelerating structure after second passage at an injection current of 0.8~mA:
		(a) --- at energy of 6~MeV,
		(b) --- at energy of 20~MeV.
	}
\end{figure}

The degree of nonlinearity of the longitudinal phase space for an injection current of 0.8~mA increases, as can be seen from Fig.~\ref{fig:electron_mode__long_phase_spaces}~(a), respectively, the total phase length of the bunch after passing through the mirror reaches almost $40^{\circ}$ --- Fig.~\ref{fig:electron_mode__long_phase_spaces}~(b).
However, the most of particles are concentrated in narrow phase range, so the width of the energy spectrum at the exit window does not increase as compared with bremsstrahlung mode --- Fig.~\ref{fig:electron_mode__spectra}.

Given the significant current margin of the accelerated beam, to reduce the width of the energy spectrum when working in the electron mode, a collimator installed in a $270^{\circ}$  bending magnet can be used.

\section{ RF power estimation and choice of the coupling with the feeding waveguide}

The klystron pulsed RF power $P_{kl}$ required to obtain the pulsed beam power $P_b$ at the bremsstrahlung target is determined by the expression:
\begin{equation}
P_{kl}=P_b+P_{loss}+P_{coll}+P_w+P_r+P_{wg}
\label{eq:klystron_power}
\end{equation}
where $P_b=E_b I_b$, $E_b$ is the energy, $I_b$ is the pulsed current at the target, $P_{loss}$ is the losses of the beam power in the accelerating structure, $P_{coll}$ is the beam power losses at the collimators, $P_w$ is the RF power losses in the walls of the accelerating structure, $P_r$ is reflected RF power, $P_{wg}$ is the RF power losses in the waveguide and ferrite isolator.

It follows from (\ref{eq:current_energy_relationship}) that for $D= 5 \text{~Gy/min}$, $q= 840$ at energy $E_b= 6 \text{~MeV}$, the pulsed beam power is $P_b \approx 0.63 \text{~MW}$, and at $E_b= 20 \text{~MeV}$, $P_b\approx 0.24 \text{~MW}$.

The RF power losses in the accelerating structure walls does not depend on the final energy of the beam, since the accelerating structure during the first passage provides a fixed energy of 10~MeV and amounts to $P_w\approx 2.7 \text{~MW}$.

\begin{figure}[thbp]
	\begin{minipage}[t]{0.49\linewidth}
		\begin{center}
			\includegraphics[width=1\linewidth]{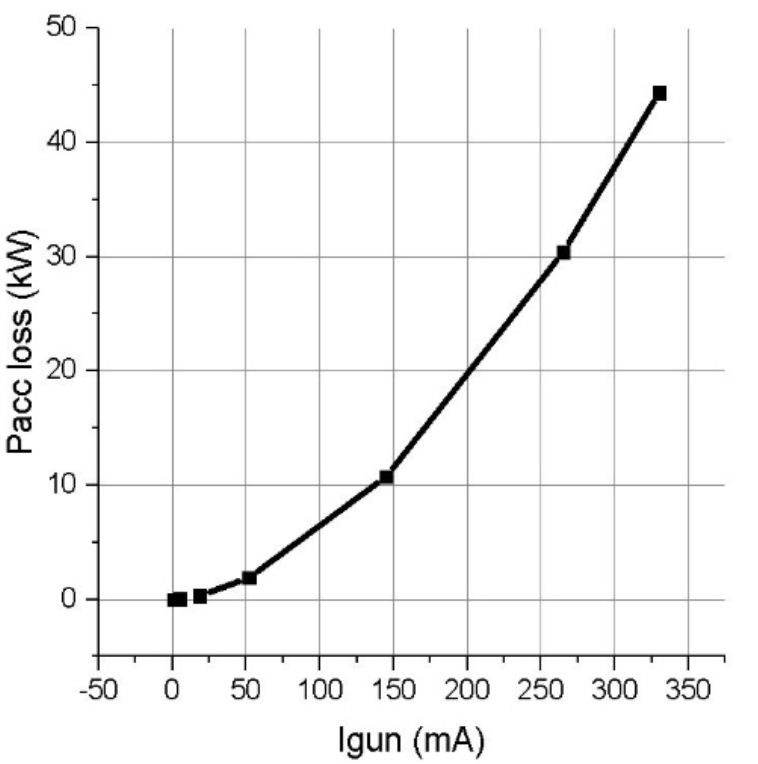}\\
			(a)
		\end{center}
	\end{minipage}
	\hfill
	\begin{minipage}[t]{0.49\linewidth}
		\begin{center}
			\includegraphics[width=1\linewidth]{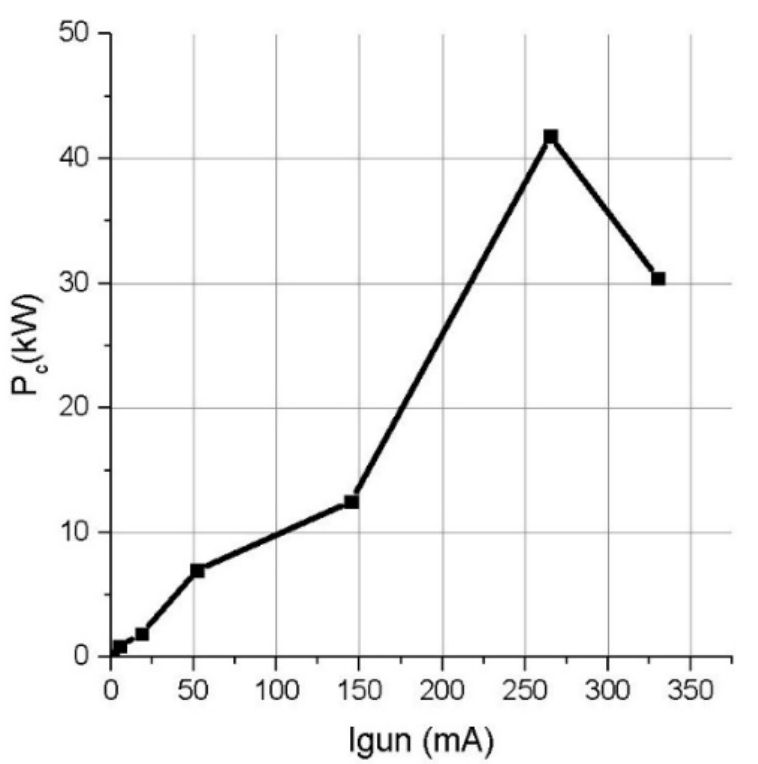}\\
			(b)
		\end{center}
	\end{minipage}
	\caption{%
		\label{fig:pulsed_beam__power_losses}%
		Pulsed beam power losses
		(a) in the accelerating structure during the first passage; 
		(b) at the collimator of a magnetic mirror.
	}
\end{figure}

Pulsed beam power losses in the accelerating structure occur mainly during the first passage of the beam and, as follows from Fig.~\ref{fig:pulsed_beam__power_losses}~(a), vary from 44~kW to 2~kW when changing the gun current from 330~mA to 50~mA.
The beam power losses at the magnetic mirror collimator (Fig.~\ref{fig:pulsed_beam__power_losses}~(b)) varies from 41~kW to 7~kW.

Thus, the total RF power consumed by the beam, $P_b^{tot}= P_b+P_{loss}+P_{coll}$, is at energy of 6~MeV about 0.7~MW, at energy of 20~MeV about 0.25~MW.

The optimal coupling coefficient of the feeding waveguide with the accelerating structure:
\begin{equation}
\beta=1+\frac{P_b^{tot}}{P_w}
\label{eq:feeding_waveguide__coupling_coefficient}
\end{equation}
is approximately $\beta \approx 1.4$ for low energy and $\beta \approx 1.1$ for high.
From the point of view of the amount of the reflected RF power, the difference in the values of the coupling coefficients is not significant, moreover, to reduce the time constant for filling the accelerating structure with electromagnetic energy, it is reasonable to choose a larger value of the coupling coefficient.
For further consideration, we put $\beta \approx 1.5$.

A peculiarity of the operation of the accelerating structure with the beam passing in the forward and backward directions is that when moving in the backward direction, depending on the phase of the bunch relative to the phase of the accelerating field, the beam can both consume electromagnetic field energy and return it.
So, for example, with the phase $\phi<-\pi / 2$ (Fig.~\ref{fig:phase}), there is a return (recovery) of energy.

In the representation of the accelerating structure as an equivalent cavity with an infinitely narrow gap, the relationship between the beam energy after the second passage with the RF power at structure input port $P_{in}$ and the beam current is determined by the expressions (\ref{eq:waveguide_relationship}), where $V_c$ is gap voltage, $\psi$ is the detuning angle, $\psi=  \tan^{-1}(2Q_L  (f_g-f)/f)$, $Q_L$, is the loaded quality factor, $f_g$ is the frequency of the RF source, $f$ is the resonant frequency of the accelerating structure, $\phi_{b1,2}$ is the beam phase with respect to $V_c$.
See e.g. \cite{Wilson__1981__High_energy_electron_linacs_application_to_storage_ring_RF_systems_and_linear_colliders}.
For further consideration, we set
$I_{b1} = I_{b2} \equiv I_b$, $\phi_{b1} = 0$, then obtain expressions (\ref{eq:waveguide_relationship_further}).

\begin{widetext}
\begin{subequations}
	\label{eq:waveguide_relationship}
		\begin{equation}
			\begin{aligned}
				V_c=
				\frac{\cos(\psi)}{1+\beta}
				\{
				&\sqrt{4\beta Z_e P_{in}-[I_{b1}Z_e \sin(\psi-\phi_{b1})+I_{b2}Z_e \sin(\psi-\phi_{b2})]^2} \\
				&-I_{b1} Z_e \cos(\psi-\phi_{b1})
				-I_{b2} Z_e \cos(\psi-\phi_{b2})
				\}
			\end{aligned}
			\label{eq:waveguide_relationship__v}
		\end{equation}
		\begin{equation}
			E_b=V_c[\cos(\phi_{b1})+\cos(\phi_{b2})]
			\label{subeq:waveguide_relationship__e}
		\end{equation}
\end{subequations}

\begin{subequations}
	\label{eq:waveguide_relationship_further}
		\begin{equation}
		\begin{aligned}
		V_c=
		\frac{\cos(\psi)}{1+\beta}
		\{
		&\sqrt{4\beta Z_e P_{in}-(I_bZ_e)^2[\sin(\psi)+\sin(\psi-\phi_{b2})]^2} \\
		&-I_b Z_e \cos(\psi)
		-I_{b} Z_e \cos(\psi-\phi_{b2})
		\}
		\end{aligned}
		\label{eq:waveguide_relationship_further__v}
		\end{equation}
		\begin{equation}
		E_b=V_c[1+\cos(\phi_{b2})]
		\label{subeq:waveguide_relationship_further__e}
		\end{equation}
\end{subequations}
\end{widetext}

\begin{figure}[thbp]
	\begin{center}
		\includegraphics[width=0.5\linewidth]{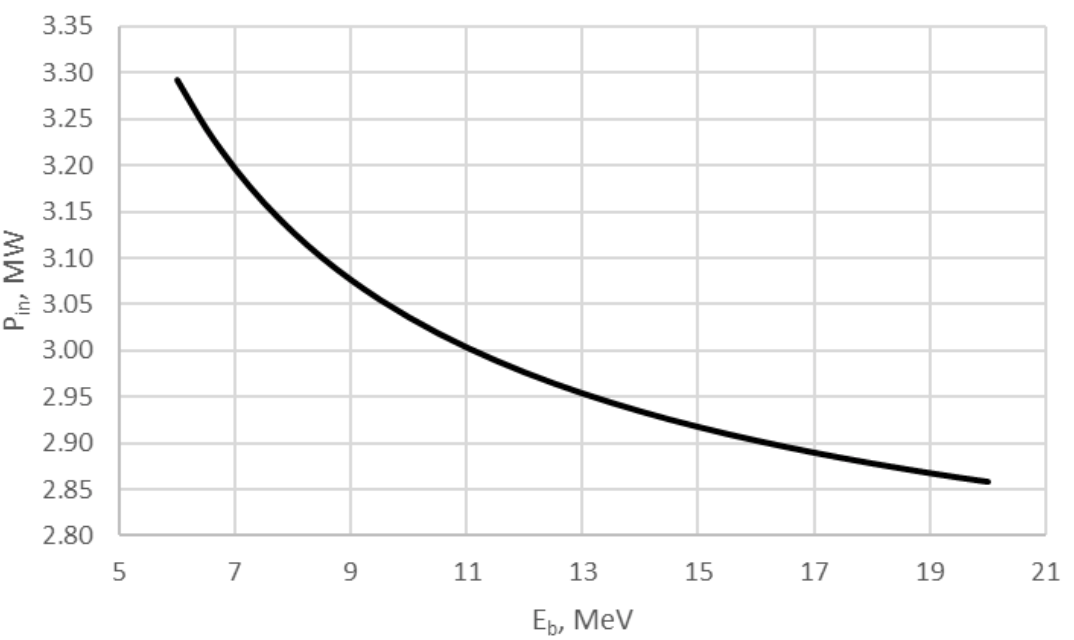}
	\end{center}
	\caption{%
		\label{fig:rf_power__vs__e}%
		The dependence of the RF power, which must be provided to the input port of the accelerating structure, on the final beam energy to produce bremsstrahlung dose rate of 5~Gy/min at a distance of 1~m from the target with a flattering filter. \\ \\ \\
	}
\end{figure}

Using expressions (\ref{eq:feeding_waveguide__coupling_coefficient}), (\ref{eq:waveguide_relationship_further}) and the values
$\psi = 0$, $V_c = 10 \text{~MV}$, $Z_e=38.5 \text{~M}\Omega$, $\beta = 1.5$, $D = 5 \text{~Gy/min}$, $q=840$, we obtain the dependence of the RF power that must be provided to the accelerating structure input port, from the final energy of the beam, presented in Fig.~\ref{fig:rf_power__vs__e}.

The maximum klystron power must exceed indicated in Fig.~\ref{fig:rf_power__vs__e} input power by the amount of losses in the waveguide path and beam power losses in the accelerating structure and at the collimator of the magnetic mirror.

\section{Conclusion}

We presented the results of calculation of the electron accelerator for external radiation therapy in the energy range of 6 -- 20~MeV.
The accelerator is based on the principle of double beam acceleration in the same accelerating structure, which allows to control the beam energy in a wide range, reduce RF power consumption and the dimensions of the accelerator, and, therefore, reduce its cost.
The results can be used to develop the design of the accelerator on the platform of the KLT-6 complex created by ROSATOM.

\bibliographystyle{nar}



\bibliography{ovchinnikova__2020__arxiv__electron_accelerator_for_radiation_therapy_with_beam_energy_6_-_20_mev}

\end{document}